\documentclass[12pt,a4paper]{article}
\usepackage{amsmath}
\usepackage{amssymb}
\usepackage[usenames,dvipsnames]{xcolor}
\usepackage{graphicx}
\usepackage{cite}
\usepackage[colorlinks=true,citecolor=OliveGreen,urlcolor=Blue,unicode,pdfhighlight=/P]{hyperref}
\usepackage[a4paper,lmargin=2cm,rmargin=1.5cm,tmargin=1.5cm,bmargin=1.5cm]{geometry}

\begin{document}
%
\title{Chaotic Lagrangian transport and mixing in the ocean}
\author{S.V. Prants}
\date{Laboratory of Nonlinear Dynamical Systems,\\
Pacific Oceanological Institute of the Russian Academy of Sciences,\\
43 Baltiiskaya st., 690041 Vladivostok, Russia, \\[1pt]
prants@poi.dvo.ru, URL: \url{http://dynalab.poi.dvo.ru}}
\maketitle
\abstract{ Dynamical systems theory approach has been successfully
used in physical oceanography for the last two decades to study
mixing and transport of water masses in the ocean. The basic
theoretical ideas have been borrowed from the phenomenon of
chaotic advection in fluids, an analogue of dynamical Hamiltonian
chaos in mechanics. The starting point for analysis is a velocity
field obtained by this or that way. Being motivated by successful
applications of that approach to simplified analytic models of
geophysical fluid flows, researchers now work with satellite-derived
velocity fields and outputs of sophisticated numerical models of
ocean circulation. This review article gives an introduction to some
of the basic concepts and methods used to study chaotic mixing and
transport in the ocean and a brief overview of recent results with
some practical applications of Lagrangian tools to monitor spreading
of Fukushima-derived radionuclides in the ocean.} 

\section{Chaotic advection in fluids: from lab to geophysical flows}
\label{sec:1}

It is well known that dynamical chaos may occur in
simple deterministic mechanical systems. One-dimensional physical pendulum under the influence
of a periodic force can move strictly periodically under some initial conditions and
is able to rotate irregularly under another ones. Roughly speaking, dynamical chaos
means that a distance between initially
nearby trajectories in the phase space grows exponentially in time
\begin{equation}
\| \delta \vec r(t) \| = \| \delta \vec r(0) \| e^{\lambda t},
\label{Lyap}
\end{equation}
where $\lambda$ is a positive number, known as the maximal Lyapunov exponent,
which characterizes asymptotically (at $t\to \infty$) the average rate
of that separation, and $\|\cdot\|$ is a norm of the position vector
$\vec r$. It immediately follows from (\ref{Lyap})
that nobody able to forecast the pendulum position $x$
beyond the so-called predictability horizon
\begin{equation}
T_{\rm pred}\simeq\frac{1}{\lambda}\ln\frac{\|\Delta_x\|}{\|\Delta_x (0)\|},
\label{horizon}
\end{equation}
where $\|\Delta_x\|$ is a confidence interval 
and $\|\Delta_x (0)\|$ is a practically inevitable inaccuracy in
specifying pendulum's initial position. The deterministic dynamical system
with positive maximal Lyapunov exponents for almost all
initial positions and momenta (in the sense of nonzero measure) is called fully chaotic.
The phase space of a typical chaotic Hamiltonian system contains islands of regular motion
embedded in a stochastic sea. The dependence of the predictability horizon
$T_{\rm pred}$ on the lack of our knowledge of exact location is logarithmic, i.e.,
it is much weaker than on the measure of dynamical instability quantified
by $\lambda$. Simply speaking, with any reasonable degree of accuracy on
specifying initial conditions there is a time interval beyond which the
forecast is impossible, and that time may be rather small for chaotic systems.
It is the ultimate reason why the exact weather forecast is impossible no matter
how perfect detectors for measuring initial parameters and how powerfull
computers we have got.

\subsection{What is chaotic advection}
\label{sec:1.1}

Methods of theory of dynamical systems have been actively used in the last 30 years to
describe advection of passive particles in fluid flows on a large range of scales, from microfluidic
flows to ocean and atmospheric ones. If advected particles rapidly adjust their
own velocity to that of a background flow and do not affect the flow
properties, then they are called passive and satisfy simple equations of motion
\begin{equation}
\frac {d{\mathbf r}}{dt} = {\mathbf v}({\mathbf r},t),
\label{adv}
\end{equation}
where ${\mathbf r}=(x, y, z)$ and ${\mathbf v}=(u, v, w)$ are the position
and velocity vectors at a point $(x, y, z)$. This formula just means that
the Lagrangian velocity of a passive particle (the left side of Eq.~(\ref{adv})) equals
to the Eulerian velocity of the flow at the location of that particle
(the right side of Eq.~(\ref{adv})).
In fluid mechanics by
 passive particles one means water (air) small parcels with their properties
or small foreign bodies in a flow. If the Eulerian velocity field is supposed to be regular, 
the vector
equation ~(\ref{adv}) in nontrivial cases is a set of three nonlinear deterministic
differential equations whose phase space is a physical space for
advected particles. Solutions of those equations can be chaotic in the
sense of exponential sensitivity to small variations in initial conditions
and/or control parameters as in Eq.~(\ref{Lyap}).

As to advection equations, it was Arnold \cite{A65} who firstly suggested
chaos in the field lines (and, therefore, in trajectories) for a special
class of three-dimensional stationary flows (so-called ABC flows), and this
suggestion has been confirmed numerically by H{\'e}non \cite{H66}. The term
``chaotic advection'' has been coined by Aref \cite{Ar84,Ar02} who realized that
advection equations for two-dimensional flows may have a Hamiltonian form.
For incompressible planar flows, the velocity components
can be expressed in terms of a streamfunction. The equations of
motion (\ref{adv}) have now the Hamiltonian form
\begin{equation}
\frac{dx}{dt}=u(x,y,t)=-\frac{\partial\Psi}{\partial y},\quad
\frac{dy}{dt}=v(x,y,t)=\frac{\partial\Psi}{\partial x},
\label{adv_eq}
\end{equation}
with the streamfunction $\Psi$ playing the role of a Hamiltonian.
The coordinates $(x,y)$ of a particle are canonically conjugated variables.
All time-independent one-degree-of-freedom  Hamiltonian systems are known to be
integrable. It means that all fluid particles move along streamlines of a
time-independent streamfunction in a regular way. Equations ~(\ref{adv_eq}) with a time-periodic
streamfunction are usually non-integrable, giving rise to chaotic particle's trajectories.
Chaotic advection has been studied both analytically and numerically in a number of
simple models with point vortices and in laboratory experiments~\cite{Ar02,Ottino}.

\begin{figure}[!htb]\center
\includegraphics[width=0.49\textwidth,clip]{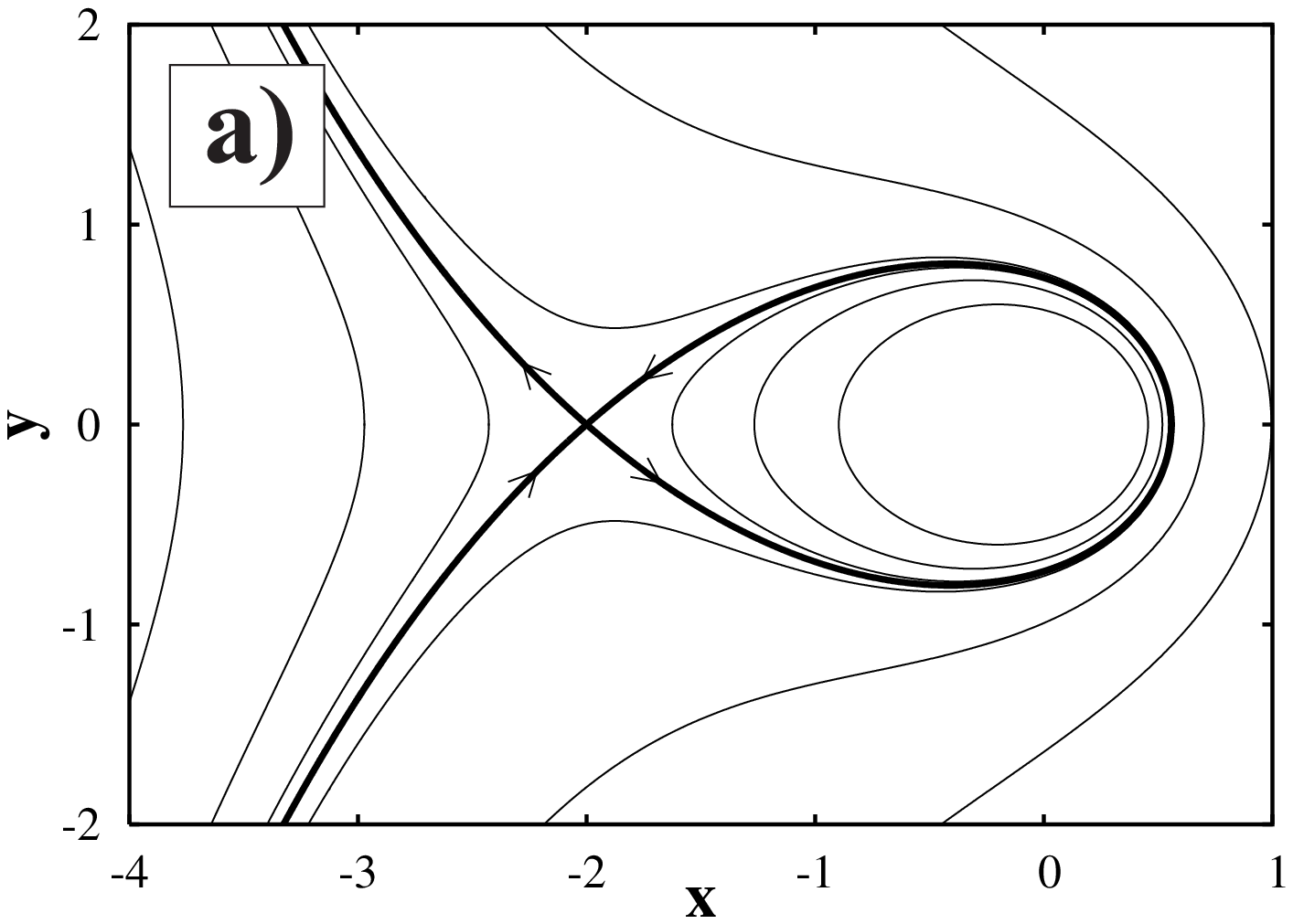}
\includegraphics[width=0.49\textwidth,clip]{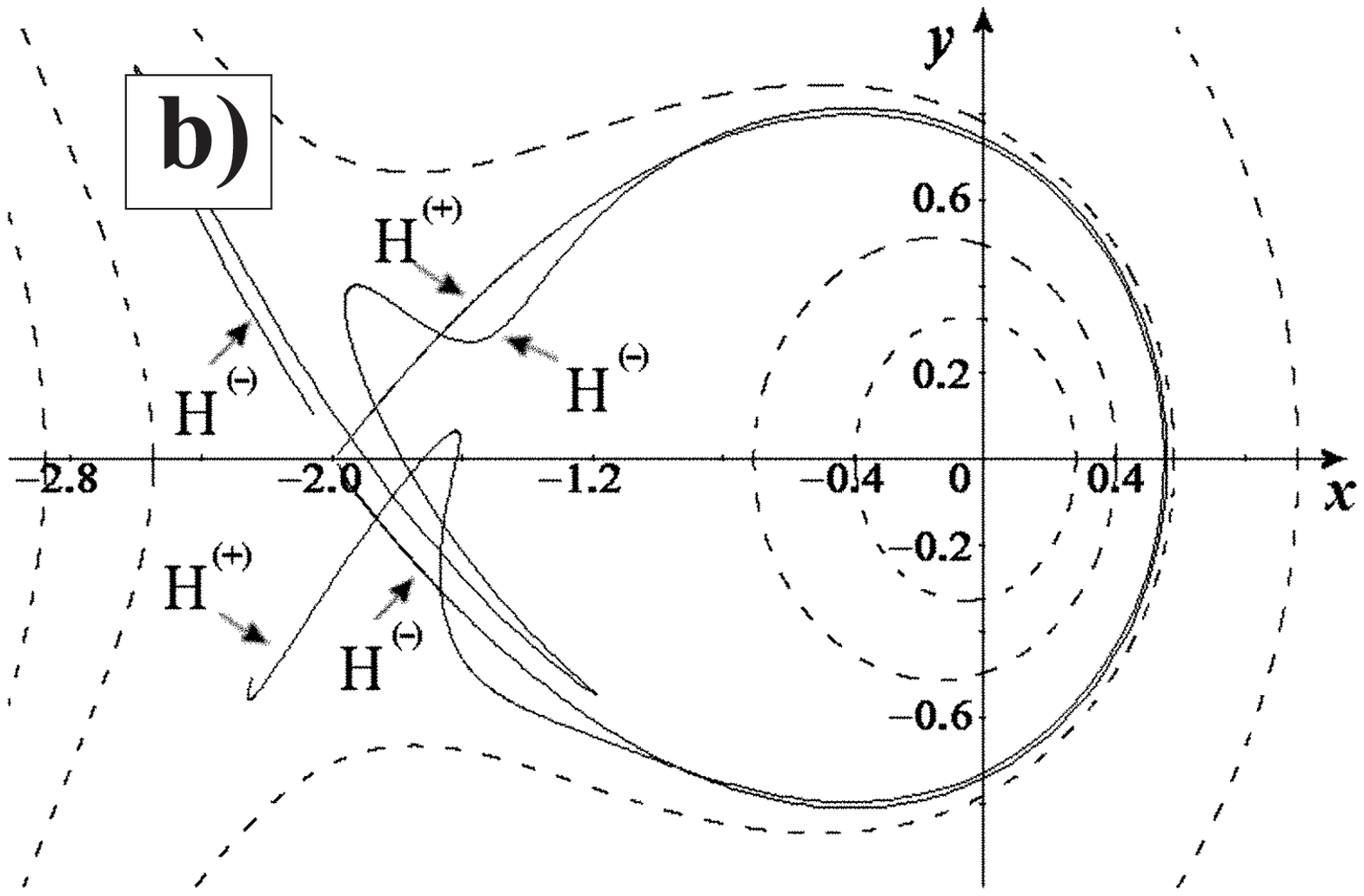}
\caption{a) A steady flow with the HSP whose stable and unstable manifolds coinside forming a closed
separatrix. b) Under a perturbation the HSP becomes a hyperbolic trajectory
with intersecting stable $H^{(+)}$ and unstable $H^{(-)}$ manifolds. The stochastic layer around
the unperturbed separatrix is a seed of Hamiltionian chaos.}
\label{fig1}
\end{figure}

Since the phase plane of the 2D dynamical system (\ref{adv_eq}) is
the physical space for fluid particles, many abstract mathematical
objects from dynamical systems theory (stagnation points, KAM tori,
stable and unstable manifolds, periodic and chaotic orbits, etc.)
have their material analogues in fluid flows. It is well known that
besides ``trivial'' elliptic stagnation points (ESP), the motion around which
is stable, there are hyperbolic stagnation  points (HSP)
which organize fluid motion in their neighborhood in a specific way.
In a steady flow the hyperbolic points are typically connected by
the separatrices which are their stable and unstable invariant
manifolds (Fig. \ref{fig1}a). In a time-periodic flow they are
replaced by the corresponding hyperbolic trajectories (HTs) with two
associated invariant manifolds which in general intersect each other
transversally (Fig. \ref{fig1}b) resulting in a complex manifold
structure known as homo- or heteroclinic tangles. The fluid motion
in these regions is so complicated that it may be strictly called
chaotic, the phenomenon known as chaotic advection. Initially close
fluid particles in such tangles rapidly diverge providing very
effective mechanism for mixing. The phase space of a typical chaotic
open flow consists of different kinds of invariant sets --- KAM tori
with regular trajectories of fluid particles, cantori and a chaotic
saddle set --- embedded into a stochastic sea with chaotic
trajectories (see, e.g., Refs.~\cite{PhysD04,JETP04} where it is illustrated with 
a simple open flow).

Stable and unstable manifolds are important organizing structures in the flow because they
attract and repel fluid particles not belonging to them and
partition the flow into regions with distinct regimes of motion.
Invariant manifold in a 2D flow is a material line, i.~e.,
it is composed of the same fluid particles in course of time.
By definition, stable,  $H^{(+)}$, and unstable, $H^{(-)}$, manifolds of a hyperbolic
trajectory $\gamma(t)$~are material lines consisting of a set of points
through which at time moment $t$ pass trajectories asymptotical to
$\gamma(t)$ at $t \to \infty$ ($H^{(+)}$) and $t \to -\infty$ ($H^{(-)}$).
They are complicated curves infinite in time and space (in theory) that act
as barriers to fluid transport (see, e.g., \cite{Wiggins05,KP06}).

\subsection{Chaotic advection in analytic models of geophysical flows}
\label{sec:1.2}

The famous paradigm of dissipative dynamical chaos, the Lorenz attractor,
appeared as a toy model of atmospheric circulation. The first
examples of Hamiltonian dynamical chaos in the ocean have been reviewed in
\cite{Zaslavsky} for describing sound propagation in so-called
underwater sound channels in the ocean where acoustic waves can
propagate over long distances practically without losses
\cite{Chaos04} (for recent reviews on ray and wave chaos in
underwater acoustics see \cite{Makarov,UFN12}).

The present article focuses on some advances in describing Lagrangian transport and mixing in the ocean
that have been made for 2D flows. The assumption that the motion is two-dimensional is justified
partly by the fact that typically horizontal velocities in the ocean are much larger that the
vertical ones, by four orders of magnitude ($10^{-1}$ vs $10^{-5}$~m/s).  2D flows are particularly
relevant for studies of pollution transport and other processes on the ocean surface. However,
2D analysis should be applied with caution if one study transport and mixing in
the areas with strong upwelling where vertical velocities may be comparable with the horizontal ones.

Kinematics of an incompressible 2D fluid flow is described by a
streamfunction $\Psi$ which gives a complete description of the
velocity field through Eq.(\ref{adv_eq}). If the streamfunction is
specified without any respect to laws of fluid motion, then the
model is said to be kinematic. The simplest analytic models of
chaotic advection in the ocean are kinematic. In fact, kinematic
streamfunctions are constructed ``by hands'' based on some heuristic
assumptions. If the streamfunction satisfies to one of the governing
equations of fluid motion, the corresponding models are called
dynamical ones. In spite of simplicity of those analytic models,
they have provided a useful insight into the problem of transport
and mixing in meandering jet currents and vortex flows. Powerful
oceanic currents, such as the Gulf Stream in the North Atlantic and
the Kuroshio in the North Pacific, are meandering jets which
transport a large amount of heat and release that to the atmosphere
strongly affecting climate. They are regions with one of the most
intense air-sea heat exchange and the highest eddy kinetic energy
level. Transport of water masses across strong jet currents, is
important because they separate waters with distinct
bio-physico-chemical properties. It may cause heating and freshing
of waters with a great impact on the weather and living organisms.

Chaotic mixing and transport in jet and vortex flows have been extensively
studied with time-periodic kinematic models (see, e.g.,
\cite{S92,Chaos06,Chaos07,JPA08,PRE09} and references therein)
and with time-periodic dynamical models conserving the potential vorticity
(see, e.g., \cite{P91,P94,DM93,Kozlov99,Rypina,Koshel08,Koshel10,PRE10,JETP10,Zhmur11,Sok11,Ryz11,Koshel13} and references therein).
The problem has been studied as well
in laboratory where azimuthal jets with Rossby waves have been
produced in rotating tanks \cite{SMS89,SHS93}.
It has been found both numerically and experimentally  that fluid
is effectively mixed along the jet, but in common opinion a large
gradient of the potential vorticity in the central part prevents
transport across the jet under realistic values
of the Rossby wave amplitudes. The transport barrier was conjectured,
based on numerical results \cite{Rypina}, to be broken only with so large amplitude values
that cannot be reached in real flows.
However, in Refs.~\cite{PRE10,JETP10} it has been shown analytically
and numerically that chaotic cross-jet transport under
appropriate conditions is possible at comparatively small values of the wave
amplitudes and, therefore, may occur in geophysical jets.
A general method has been elaborated in those papers to detect a core of the transport
barrier and find a mechanism of its destruction using the dynamical
model of a zonal jet flow with two propagating Rossby waves. The
method comprises the identification of a central invariant curve,
which is an indicator of existence of the barrier, finding certain resonance
conditions for its destruction and detection of cross-jet transport.

The real oceanic flows are not, of course, strictly time-periodic.
In aperiodic flows there exist HSPs and HTs as well, but of a transient nature.
It is possible to identify aperiodically
moving HSPs with stable and unstable effective manifolds
\cite{H02}. Unlike the manifolds in steady and periodic flows, defined
in the infinite time limit, the ``effective'' manifolds of aperiodic
HSPs  have a finite lifetime. The point is that they
play the same role in organizing oceanic flows as do invariant manifolds
in simpler flows. The effective manifolds in course of their life undergo
stretching and folding at progressively small scales and intersect each other
at homoclinic points in the vicinity of which fluid particles move
irregularly. Trajectories of initially close fluid particles diverge rapidly
in those regions, and particles from other regions appear there. It is the
mechanism for effective transport and mixing of water masses in the ocean.
Moreover, stable and unstable effective manifolds constitute Lagrangian
transport barriers between different regions, because they are material
invariant curves that cannot be crossed by purely advective processes.

The stable and unstable manifolds of influential HSPs
are so important because (1) they form a kind of a skeleton in oceanic flows,
(2) they partition a flow in dynamically distinct regions,
(3) they provide inhomogeneous mixing with
spirals, filaments and intrusions which are often visible on satellite images, 
(4) they are transport barriers
separating water masses with different bio-physico-chemical characteristics.
Stable manifolds act as repellers for surrounding waters
but unstable ones are a kind of attractors.
That is why unstable manifolds may be rich in nutrients being oceanic ``dining rooms''.

\subsection{Lagrangian coherent structures}
\label{sec:1.3}

The existence of large-scale quasi-deterministic coherent structures
in quasi-random (turbulent) flows has long been recognized (see, e.g., \cite{Hussain83}).
Before the coherent structures were found, it was a common opinion that
turbulent flows are determined only by irregular vortical fluid motion.
Although up to now there is no consensus on a strict definition of coherent structures,
they can be considered as connected turbulent fluid masses
with phase-correlated (i.e., coherent) vorticity over the spatial extent of the shear layer.
Thus, turbulence consists of coherent and phase-random (incoherent) motions with the latter ones to be
superimposed on the former ones. Lagrangian motion
may be strongly influenced by those coherent structures that support distinct regimes in
a given turbulent flow. The discovery that
turbulent flows are not fully random but embody orderly organizing structures was
a kind of revolution in fluid mechanics.

As to complicated but not totally random flows, including
large-scale geophysical ones, it was Haller \cite{H02} who proposed
a concept of Lagrangian coherent structures (LCS) with the
boundaries delineated by distinguished material lines (surfaces) and
advected with the flow. To extract these structures, he proposed to
compute finite-time Lyapunov exponents
(FTLE). The LCS are operationally defined as local extrema of the
scalar FTLE field, $\lambda (x,y)$, which characterizes the 
rate of the fluid particle dispersion over a finite-time interval.
They are the most influential attracting and repelling hyperbolic
material curves in 2D velocity fields. The LCS are Lagrangian
because they are invariant material curves consisting of the same
fluid particles. They are coherent because they are comparatively
long lived and more robust than the other adjacent structures.  The
LCS are connected with stable and unstable invariant manifolds of
HSPs. A tracer patch, chosen nearby any stable manifold,
moves in the course of time to the corresponding HSP squeezing along
that manifold. After approaching the HSPs, the patch begins to stretch along
the corresponding unstable manifold.

The LCS are supposed to be the most repelling and attracting material lines in 2D flows \cite{H02}.
The spatial distribution of the FTLE values backward and forward in time 
is an effective way to compute them. A region under study
is seeded with a large number of tracers on a grid. The FTLE values are computed by one of 
the known methods
(one of them is described in Sec.~3.2) for all neighboring grid points for a given period of time,
typically from a week to a month. Then one plots
the spatial distribution of the FTLE coding its values by color. If there were hyperbolic regions
in the velocity field
in the area for a chosen period of time, then we should get a spatially inhomogeneous FTLE map
with ``ridges''  and ``valleys''. A ``ridge'' is defined as a curve on which the FTLE
is locally maximized in the transverse direction \cite{Shadden05}. Both the repelling and attracting
LCS can be computed by this way. Integrating the advection equations forward in time and computing
the FTLE ``ridges'',
we extract repelling LCS which approximate influential stable manifolds in the area.
Expansion in backward time
implies contraction in forward time. Therefore, attracting LCS can be computed analogously but
in reverse time. They approximate influential unstable manifolds in the area.

A hard work has been done by many people to enlarge the notion of the LCS and invariant manifolds
to finite-time realistic flows. In numerous papers (see, e.g., recent introductory review papers
\cite{Haller13,Samelson13}, papers
\cite{Abraham02,Ovidio04,Shadden05,Kirwan06,Lehahn07,Beron08,Kai09,OM11,DAN11,Huhn12,DAN12,Bettencourt12,FAO13,P13,Beron13}
and references therein) the LCS have been shown to
be very useful mean to analyze mixing and transport in different seas and oceans.
The aim of this paper is not to focus on the LCS but rather to focus on
recent studies of transport and mixing in the ocean using Lagrangian indicators 
which have been introduced to physical oceanography only recently.


\section{Lagrangian approach to study transport and mixing in the ocean}
\label{sec:2}

\subsection{Transport and mixing in fluids}
\label{sec:2.1}

There are two common approaches in hydrodynamics to study fluid motion,
the Eulerian and Lagrangian ones. In the Eulerian approach one is interested in
velocities of the flow at given points on a spatial grid. In the Lagrangian one
we look for trajectories of water parcels advected by an Eulerian velocity field
(\ref{adv}). It is a more convenient approach to study transport and
mixing in the ocean, especially the fate and origin of water masses.
The velocity field ${\mathbf v}({\mathbf r}, t)$ is supposed to be known
analytically, numerically or estimated from satellite altimetry data.
While in the Eulerian approach we get frozen snapshots
of data, Lagrangian diagnostics enable to quantify spatio-time variability
of the velocity field. In this review we deal with large-scale horizontal Lagrangian transport
and mixing in the ocean.
We are interested in water transport on comparatively large scales exceeding 10~km.
Transport is advection of the water mass with its conserved properties
due to the fluid's bulk motion. Advection requires currents and eddies which are vortical currents.

Mixing is a key concept both in hydrodynamics and in dynamical systems theory which can be defined
in a strict mathematical sense. Let us consider the basin $A$ with a circulation where there is
a domain $B$ with a dye occupying at $t=0$ the volume $V(B_0)$. Let us consider a domain $C$
in $A$.
The volume of the dye in the domain $C$ at time $t$ is $V(B_t \cap C)$, and its
concentration in $C$ is given
by the ratio $V(B_t \cap C)/V(C)$. The definition of full mixing is that in the course of time
in any domain $C \in A$ we will have
the same dye concentration as for the entire domain $A$, i.e.,
$V(B_t \cap C)/V(C) - V(B_0)/V(A) \to 0$ as $t \to \infty$.
In dynamical systems theory the full or global mixing is achieved when
a small blob of the phase-space fluid is transformed into a long intricate filament occupying
all energetically accessible domain in
the phase space. The mixing measures are the Lyapunov exponents. In real flows mixing due to flow
kinematics is accompanied by molecular diffusion and small-scale turbulence, the processes which
can be ignored in large-scale geophysical flows.

Chaotic advection in theory is chaotic mixing in a regular velocity field.
In real flows there are inevitable random fluctuations of that field. If they are small
as compared to mean regular values, it is reasonably to call the corresponding phenomenon as
chaotic advection, because typical geometric structures have similar forms as in
purely deterministic flows but become just more fuzzy \cite{PRE06}.
Both turbulence and chaotic advection lead to mixing. What is the difference?
Chaotic mixing may occur if the velocity field is quasi-coherent in space and quasi-regular
in time, but the motion of tracers is irregular on much more smaller scales.
In the ocean chaotic mixing produces smooth large-scale features visible sometimes on satellite images
of the ocean color as stretched and folded curves. Turbulent mixing may occur if the velocity field
is incoherent in space and irregular in time on the same scales as the motion of tracers.
It is homogeneous on comparatively large scales.

\subsection{How to get a velocity field on the sea surface}
\label{sec:2.2}

%
%

The impressive progress in the past two decades in satellite
monitoring and development of high-resolution numerical models of
ocean circulation have opened up new opportunities in physical
oceanography. At the websites \url{http://www.aviso.oceanobs.com},
\url{http://oceancolor.gsfc.nasa.gov}, \url{http://www.nodc.noaa.gov} and others
one can monitor day by day the sea surface temperature and salinity,
concentration of chlorophyll-$\alpha$, winds at the ocean surface,
sea surface height and many other things. Drifter and buoy
observations now cover most areas of the world's oceans at
sufficient density to map mean currents at one degree resolution
(\url{http://www.aoml.noaa.gov} and \url{http://www.nodc.noaa.gov}). Systematic
and routine satellite measurements of the World Ocean and atmosphere
and their rapid interpretation with the help of numerical
forecasting models provide not only final products on the present
state of the sea but continuous forecasts of its future conditions
as well.

The launch of Earth-observing altimeter satellites in the 1990s
opened a new era for studying ocean surface circulation.
A satellite radar measures precisely the distance from the radar antenna to
the ocean surface by computing the round-trip travel time of a microwave signal.
Dynamic topography refers to the topography of the sea surface related to the dynamics of its own flow.
In hydrostatic equilibrium, the surface of the
ocean would have no topography, but due the ocean currents, its maximum dynamic topography is on
the order of two
meters and are influenced by ocean circulation, temperature and salinity.
A clockwise rotation (anticyclone) is found around elevations on the ocean surface
in the northern hemisphere and depressions in the southern hemisphere.
Conversely, a counterclockwise rotation (cyclone) is found around depressions 
in the northern hemisphere and elevations in the southern hemisphere.
Combined with precise satellite location data, altimetry measurements
yield sea-surface heights which, in turn,
allow to infer ocean currents under conditions of the geostrophic balance.
Away from the surface and bottom layers, horizontal pressure gradients in the ocean
almost exactly balance the Coriolis force. The resulting flow is known as geostrophic.
The major currents, such as the Gulf Stream, the Kuroshio and the Antarctic Circumpolar Current,
are examples of geostrophic currents. Given a streamfunction $\Psi(x,y,t)=gh/f$,
one gets from (\ref{adv_eq}) the formula connecting surface geostrophic velocities with surface slope
\begin{equation}
u_{gs}=-\frac{g}{f}\frac{\partial h}{\partial y},
\qquad v_{gs}=\frac{g}{f}\frac{\partial h}{\partial x},
\label{geos}
\end{equation}
where $g$ is gravity, $f=2\Omega \sin \phi$ is the Coriolis parameter, $\Omega$ is 
the angular speed of the Earth, $\phi$ is the latitude and $h$ is the sea height above a level surface.
Daily geostrophic velocities for the world's oceans, provided by the AVISO database
(\url{http://www.aviso.oceanobs.com}), approximate geostrophic ocean currents
for horizontal distances exceeding a few tens of kilometers, and for times greater than a few days.
The velocity data covers the period from 1992 to the present time with daily data on
a $1/3^{\circ}$ Mercator grid.

Satellites can observe some processes in the ocean almost everywhere
but near the surface only. Research vessels can measure more
variables in the depth, but they are too sparse. Numerical models
seem to be a useful complementary tool to get a detailed view of the
ocean circulation. Realistic velocity fields can be obtained from
numerical multi-layered models of regional and global ocean
circulation which are able to reproduce adequately many of mesoscale
and submesoscale characteristic features of circulation including
eddies of various sizes. Those models assimilate satellite
sea-surface height and sea-surface temperature and are forced by
surface winds and air-sea fluxes to be estimated from the
Reanalysis. They provide the velocities at different depths, from a
surface layer to the bottom one. Numerical models are not free, of
course, of many sources of errors. So, one should accept their
outputs with a caution.

\section{Numerical computation}
\label{sec:3}

\subsection{Options for advection equations}
\label{sec:3.1}

The satellite-derived and numerically generated velocity fields are given as discrete data sets,
rather than analytical functions. Moreover, the velocity field in the ocean is only known for
finite times. Some numerical algorithms are needed to solve advection equations with such data sets.

Let the velocity field, $(u_{i,j},v_{i,j})$, is given on a 2D grid $(\lambda_{i,j},\varphi_{i,j})$, $i=0,1,\dots,N_x$,
$j=0,1,\dots,N_y$, where $\lambda_{i,j}$~is the longitude and $\varphi_{i,j}$~the latitude.
We suppose that the grid can be transformed to a rectangular form, i.e., there exist functions,
$X(\lambda,\varphi)$ and $Y(\lambda,\varphi)$, such that $x_{i,j}=X(\lambda_{i,j},\varphi_{i,j})=i\Delta x$ and
$y_{i,j}=Y(\lambda_{i,j},\varphi_{i,j})=j\Delta y$, where $\Delta x$ and $\Delta y$~are constants.
Simply speaking, there exists a coordinate frame where our grid is rectangular.
In the simplest case $(\lambda_{i,j},\varphi_{i,j})$ is already rectangular. However, it is 
inconvenient for global
fields due to singularities at poles and dependence of the resolution on latitude.
The velocities of oceanic
currents are usually given in the units of km per day: $1\,\text{cm/s}=0.864\text{km/day}$.

There are a few options to write down the advection equations.

1) {\em  Neither velocities nor coordinates are transformed:}
\begin{equation}
\dot\lambda=\frac{10800\cos{\varphi}}{\pi R}u,\quad \dot\varphi=\frac{10800}{\pi R}v,
\label{1eq}
\end{equation}
where $R$~the Earth's radius, latitude and longitude are in geographic minutes, time is in days and
$u$ and $v$ are in km day$^{-1}$. The coefficient $\pi R/10800$, is equal approximately
to a nautical mile $1.852$ km.
{\em Advantages:} No coordinate transformations, and inputs and outputs are given in geographic
coordinates. {\em Disadvantages:} The calculation of cosine at each step is a time-consuming procedure.
Computation of the cell with current coordinates, which one needs for interpolation,
is not simple with a nonrectangular grid.

2) {\em Transformation of linear velocities to angular ones in the right side of advection equations:}
\begin{equation}
\dot\lambda=u,\quad \dot\varphi=v,
\label{2eq}
\end{equation}
where $u$ and $v$ are in minutes per second. {\em Advantages:} The same as in the first method.
{\em Disadvantages:} The same as with nonrectangular grids.

3) {\em Transformation of coordinates and velocities to a rectangular grid:}
\begin{equation}
\dot x=u,\quad \dot y=v.
\label{maineq}
\end{equation}
{\em Advantages:} Simple and fast numerical integration. {\em
Disadvantages:} Since we compute in abstract coordinates, it is
necessary to transform input and output data.

In the oceanographic examples followed we have used Eqs. \ref{maineq}
because of a simple, unified code that does not depend on the real grid
$(\lambda_{ij},\varphi_{ij})$.

Coordinates $x$ and $y$ of a passive particle
are related with its latitude $\phi$ and longitude $\lambda$
in degrees as follows:
\begin{equation}
\lambda=\frac{x}{60},\quad
\phi=\frac{180}{\pi}\arcsin{\tanh{\left(\frac{\pi}{180}\left(\frac{y}{60}
+y_0\right)\right)}},\quad
 y_0=\frac{180}{\pi}\operatorname{artanh}{\sin{\left(\frac{\pi}{180}\phi_0\right)}},
\label{coordinate}
\end{equation}
where $\phi_0=-82$.
We use the transformation (\ref{coordinate}) because the AVISO grid is homogeneous in
those coordinates. The velocities $u$ and $v$ in eq.~(\ref{maineq}) are expressed via
latitudinal $U_\phi$ and longitudinal $U_\lambda$ components of the linear velocity
$U$ in cm/s as follows:
\begin{equation}
u=\frac{10800}{\pi R_E\cos\phi}\frac{86400}{100000}U_\lambda\approx
\frac{0.466}{\cos\phi}U_\lambda,\quad
v=\frac{10800}{\pi R_E\cos\phi}\frac{86400}{100000}U_\phi\approx \frac{0.466}{\cos\phi}U_\phi.
\label{velocity}
\end{equation}

The velocity field is given on a grid $(x_{i,j},y_{i,j},t_k)$. In order to integrate the advection equations,
we need to know velocities between the grid points interpolating a given data in space and time.
Thus, our numerical algorithm for solving advection equations is as follows.
\begin{enumerate}
\item  Transformation of coordinates and velocities to a rectangular grid with creating a file with
information about the grid
(the number and size of steps in space and time) and velocities.
\item  A bicubical interpolation in space and an interpolation by third order Lagrangian polynomials
in time are used. The velocity components are interpolated independently on each other.
\item The velocities obtained are substituted in Eq. (\ref{maineq}) which is integrated with a fourth-order
Runge-Kutta scheme with a fixed time step.
\item The outputs are analyzed and then transformed in the geographical coordinates to get images
and maps.
\end{enumerate}

\subsection{Computing finite-time Lyapunov exponents}
\label{sec:3.2}

A general method for computing finite-time Lyapunov exponents (FTLE),
which is valid for $n$-dimensional vector fields, has been proposed recently
~\cite{OM11}. The Lyapunov exponents in this method are computed via singular
values of the evolution matrix, $\sigma_i$, which obeys the differential matrix equation
\begin{equation}
\dot G=JG,
\label{evol_mat_diffur}
\end{equation}
with the initial condition $G(t_0,t_0)=I$, where $I$ is the unit matrix. Here $J$ is Jacobian matrix for
the linearized $n$-dimensional equations of motion. After a singular-value decomposition  
of the evolution matrix, one gets
the expression for the Lyapunov exponents
\begin{equation}
\lambda_i=\lim_{t\to\infty}\frac{\ln\sigma_i(t,t_0)}{t-t_0}, \quad n=1,2, ... .
\label{lyap_def_sigma}
\end{equation}
Quantities
\begin{equation}
\lambda_i(t,t_0)=\frac{\ln\sigma_i(t,t_0)}{t-t_0}
\label{lyap_ftle}
\end{equation}
are called FTLE which of each is the ratio of the logarithm of the maximal possible
stretching in a given direction to a finite time interval $t-t_0$.

The formulae above are valid with any $n$-dimensional set of nonlinear differential equations.
In the two-dimensional case of particle's advection on the ocean surface  (\ref{adv_eq})
the singular-value decomposition of the $2\times 2$ evolution matrix is
as follows:
\begin{equation}
G \equiv
\begin{pmatrix}
a&b\\c&d
\end{pmatrix}
=
\begin{pmatrix}
\cos\phi_2&-\sin\phi_2\\
\sin\phi_2&\cos\phi_2
\end{pmatrix}
\begin{pmatrix}
\sigma_1&0\\
0&\sigma_2
\end{pmatrix}
\begin{pmatrix}
\cos\phi_1&-\sin\phi_1\\
\sin\phi_1&\cos\phi_1
\end{pmatrix}.
\label{SVD2x2}
\end{equation}
Solution of these four algebraic equations are
\begin{equation}
\begin{gathered}
\sigma_1=\frac{\sqrt{(a+d)^2+(c-b)^2}+\sqrt{(a-d)^2+(b+c)^2}}{2},\\
\sigma_2=\frac{\sqrt{(a+d)^2+(c-b)^2}-\sqrt{(a-d)^2+(b+c)^2}}{2},\\
\phi_1=\frac{\operatorname{arctan2}{(c-b,\,a+d)}-\operatorname{arctan2}{(c+b,\,a-d)}}{2},\\
\phi_2=\frac{\operatorname{arctan2}{(c-b,\,a+d)}+\operatorname{arctan2}{(c+b,\,a-d})}{2},
\end{gathered}
\label{SVD2x2finsol}
\end{equation}
where function $\operatorname{arctan2}$ is defined as
\begin{equation}
\operatorname{arctan2}{(y,x)}=\left\{
\begin{aligned}
&\arctan{(y/x)}, &x\ge 0,\\
&\arctan{(y/x)}+\pi, &x<0.
\end{aligned}\right.
\label{arctg2}
\end{equation}

The method described has been applied to study large-scale transport and mixing
in different regions in the ocean \cite{OM11,DAN12,FAO13,P13,DSR14,FAO14,NPG14,Lobanov,OD14}.

\section{Elliptic and hyperbolic regions in the ocean}
\label{sec:4}

The AVISO altimetric field is provided with a day interval.
``Instantaneous'' stagnation points in such a field are those points
in a fixed day where the AVISO velocity is found to be zero. Their
local stability properties are characterized by eigenvalues of the
Jacobian matrix of the velocity field. For 2D flows, if the two
eigenvalues are real and of opposite sign, then the stagnation point
is a HSP. If they are pure imagine and complex conjugated, then one
gets an ESP. The two zero eigenvalues of the Jacobian matrix means
the existence of a parabolic stagnation point. The stagnation
points are typically moving Eulerian features in a frozen-time
velocity field. They are not fluid particle trajectories.
\begin{figure}[!htb]
\centerline{\includegraphics[width=0.5\textwidth,clip]{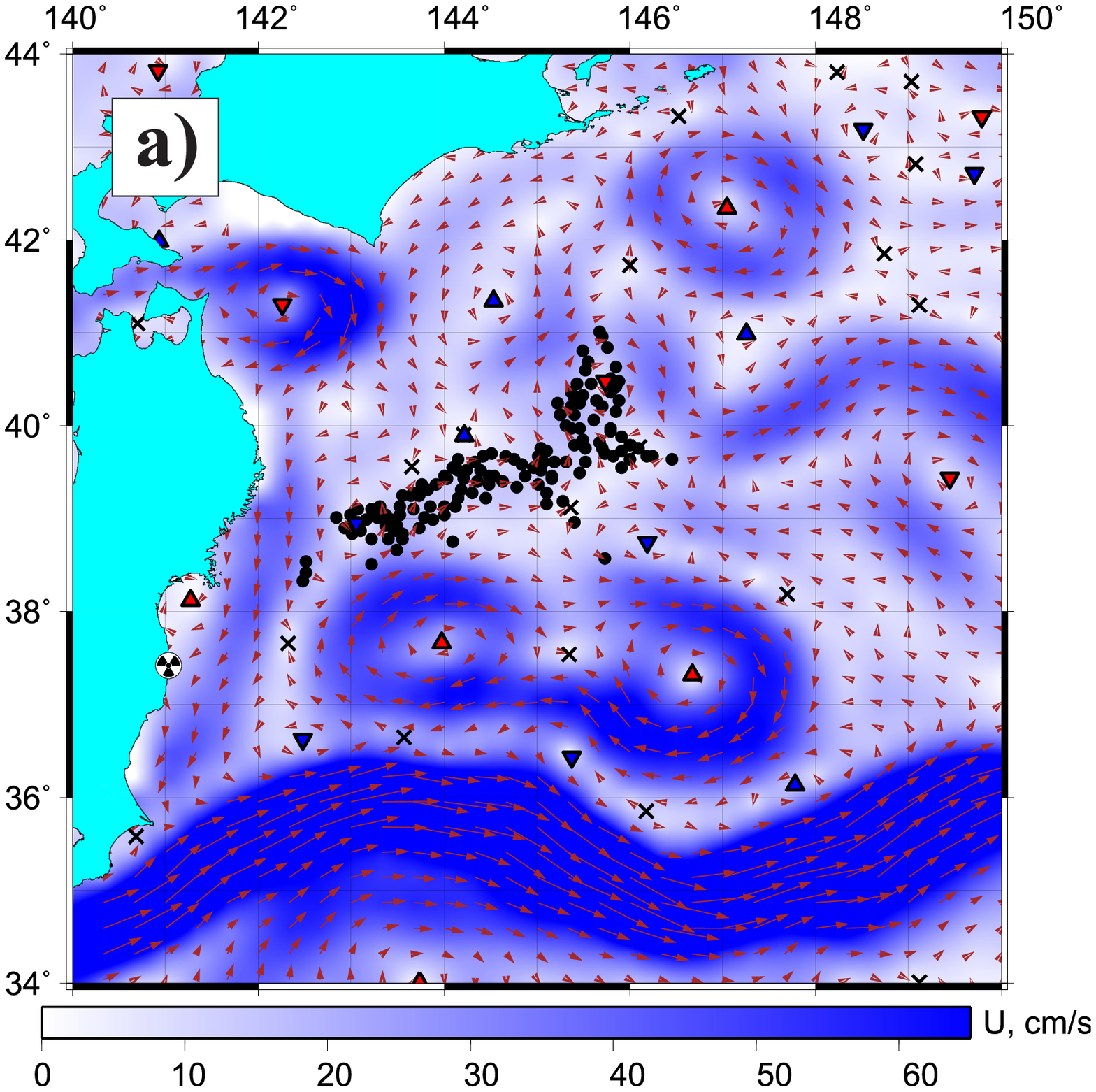}
\includegraphics[width=0.47\textwidth,clip]{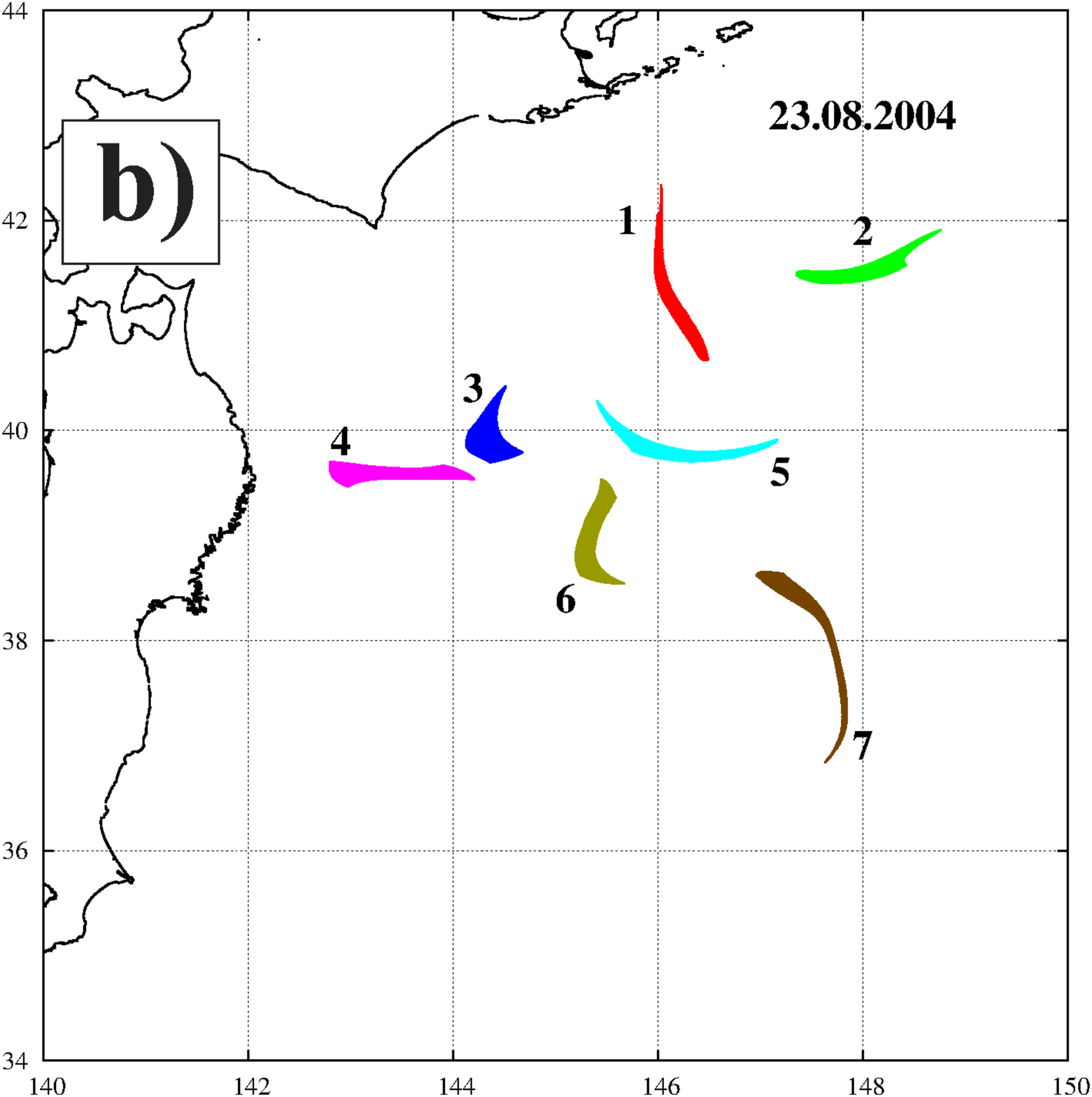}}
\centerline{\includegraphics[width=0.47\textwidth,clip]{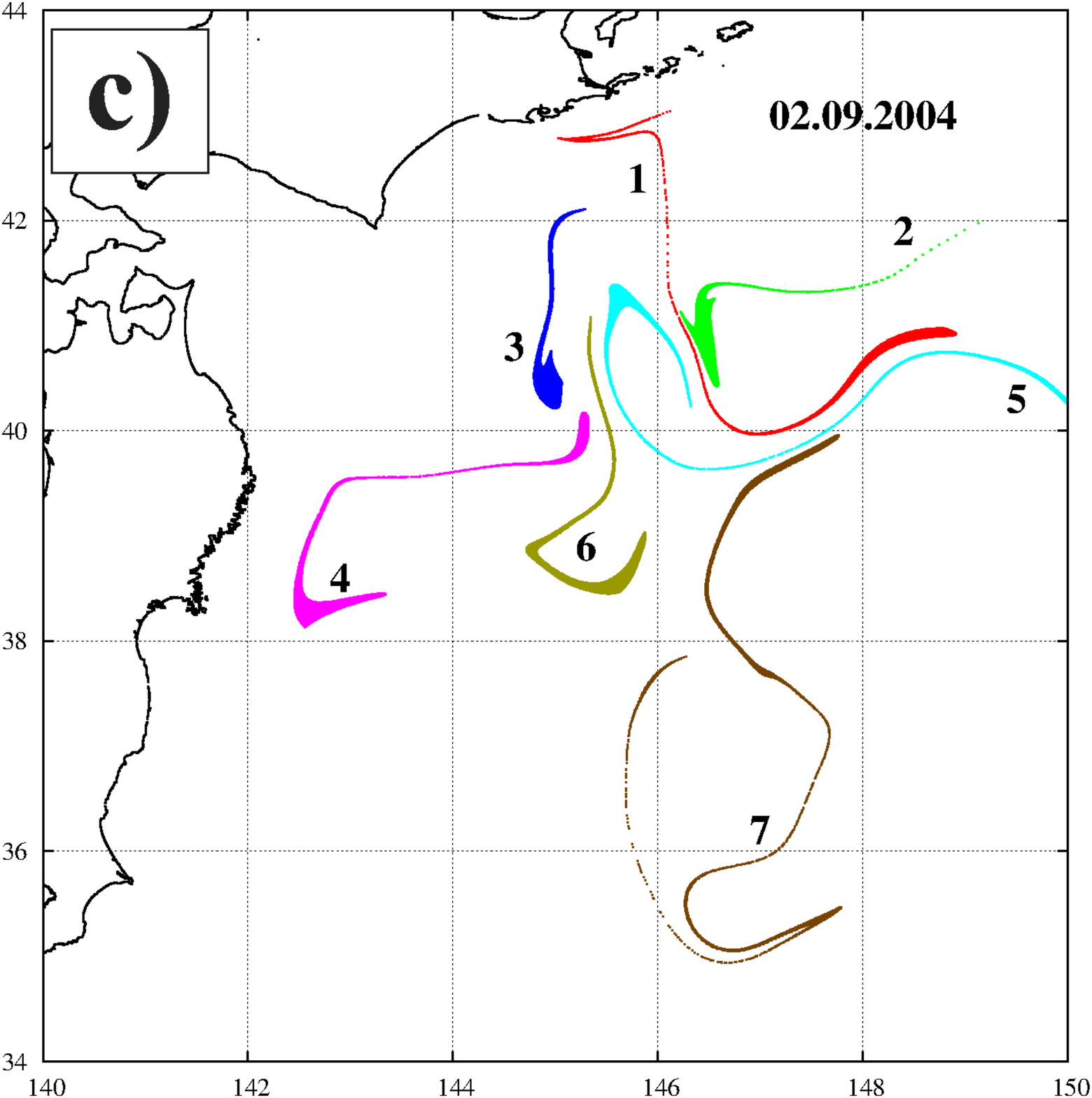}
\includegraphics[width=0.47\textwidth,clip]{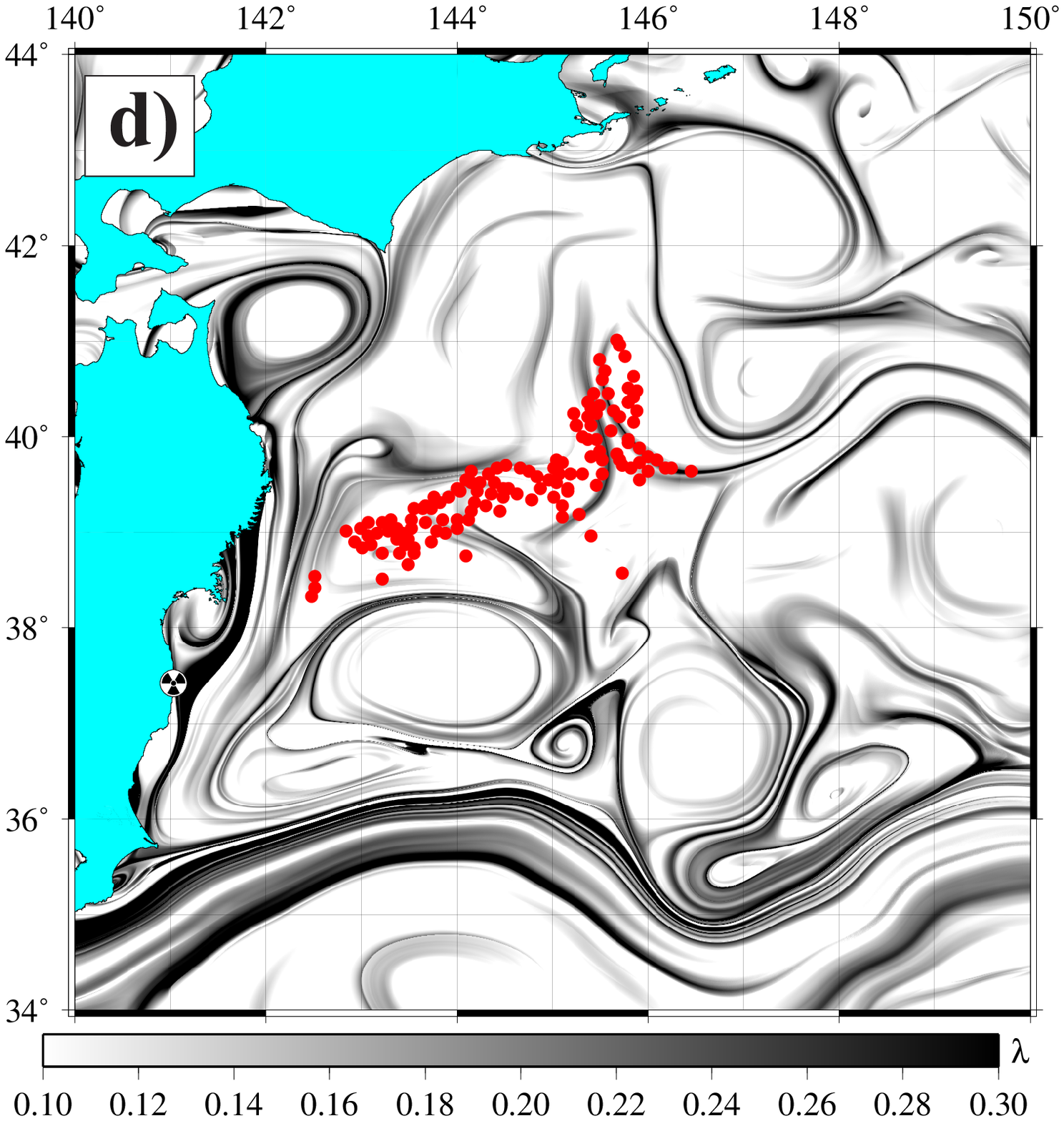}}
\caption{a) The altimetric velocity field in the North Western Pacific 
on 19 August 2004 with overlaid tuna fishing locations (dots) and elliptic 
(triangles) and hyperbolic (crosses) ``instantaneous'' stagnation points. 
b) and c) Evolution of 7
patches with synthetic particles placed on August, 19, 2004 at 7
HSPs in the region. For two weeks the patches delineate the
corresponding unstable manifolds seen as the black ridges on the
backward-time FTLE map in d) with $\lambda$ in days$^{-1}$.}
\label{fig2}
\end{figure}
The example of the altimetric velocity field in the North Western Pacific is shown
in Fig. \ref{fig2}~a
with overlaid positions of hyperbolic (crosses) and elliptic (triangles) stagnation points.
The ESPs are situated mainly in the centers of eddies and allow to identify the latters.
The saddle-type HSPs are situated mainly between the eddies or nearby single eddies.
They are important because
they are features attracting water in one direction and expelling water in the other one.

Stagnation points may undergo bifurcations. Bifurcation theory, among other things, is interested
in behavior of
fixed points of vector fields as a parameter is varied. In our case time plays the role of
the parameter.
One may monitor positions of HSPs and ESPs day by day and look for their movement around
in the  ocean
flow. Nothing interesting, besides a rearrangement of the flow, occurs if they do not change their
stability type.
When they do that, there are, in principle, a few possibilities \cite{Wig06}. In the saddle-node
bifurcation two stagnation points,
a HSP and a ESP, collide and annihilate each other in the course of time. The opposite process
could occur as well: two stagnation points, one  HSP
and one ESP, are born suddenly. After the collision stagnation points
may move apart without changing in number (transcritical bifurcation) or split into three ones
(pitchfork bifurcation).
In fact, we have observed only saddle-node bifurcations in the altimetric velocity field
in the region shown in Fig. \ref{fig2}.

Stable and unstable manifolds of hyperbolic objects can be
identified by local and global methods. In the local approach, first
of all, one locates positions of HSPs. Then it is necessary to
identify the HTs which are situated, as a rule, nearby the HSPs. It
can be done by different ways. We prefer to use a HSP as the first
guess, placing in a fixed day a few material segments oriented at
different angles and computing the FTLE for the particles on those
segments. Coordinates of the particles with the maximal FTLE give us
approximate position of the HT nearest to that HSP on that day. Then
we place the patch with a large number of synthetic particles,
centered at the HT position, and evolve it forward in time. It is
shown in Figs.~\ref{fig2}b and c how the method works in the
Oyashio -- Kuroshio frontal zone in the northwestern part of the
Pacific Ocean to the east off Japan, where the subarctic waters of the cold
Oyashio Current encounter the subtropical waters of the warm Kuroshio
Current. This region is known to be one of the richest fishery in
the world. The jet of the Kuroshio Extention in the south, the two
anticyclonic mesoscale eddies to the north of the jet, the Kuroshio
ring near the Hokkaido Island with the center at ($147^{\circ}E;
42^{\circ}.4 N$) and the anticyclonic mesoscale eddy at the traverse
of the Tsugaru straight ($142^{\circ}.3E; 41^{\circ}.4 N$) are
clearly seen in Fig.~\ref{fig2}a. Figures~\ref{fig2}b and c show how
the corresponding unstable manifolds evolve from 7 tracer patches
placed near the 7 HTs.

In the global approach, one seeds the whole area with a large number
of synthetic particles and compute the FTLE field which is a
commonly used measure of hyperbolicity in oceanic and atmospheric
flows. It has been shown by Haller \cite{H02,Haller13} that the
curves of local maxima of the FTLE field attributed to initial
tracer's positions approximate stable manifolds when computing
advection equations forward in time and unstable ones when computing
them backward in time (see Sec.~1.3). To compare the results obtained in the local
and global approaches, we compute the FTLE field in the same region.
Comparing Fig.~\ref{fig2}~c with the backward-time FTLE map on 2
September in Fig.~\ref{fig2}~d, it is seen that the patches
delineate the corresponding ridges on the backward-time FTLE map.
The patches nos. 1--5 were chosen in productive waters with
comparatively high chlorophyll-$\alpha$ concentration, but in the
course of time they have been transformed into narrow filaments to
be penetrated into oligotrophic waters, poor with nutrients. The
passive marine organisms in those fluid patches are advected along
with them into oligotrophic waters attracting fish and marine animals for feeding.

\section{Lagrangian maps and Lagrangian fronts}
\label{sec:5}

The Lyapunov maps provide valuable information on the LCS in oceanic flows.
Additional information about the origin, history and fate of water masses
can be obtained based on synoptic maps of the Lagrangian indicators measuring
some quantitites along a parcel trajectory.
Among them are vorticity, a distance passed by fluid particles for a given time,
absolute, $D$, meridional, $D_y$,  and zonal, $D_x$,  displacements of particles from their initial positions,
the time of residence of fluid particles in a given region, the number of their cyclonic
and anticyclonic rotations and others \cite{OM11,DAN11,FAO13,P13}.
The absolute displacement is simply the distance between the final, $(x_f,y_f)$, and
initial, $(x_0,y_0)$, positions of advected particles on the Earth sphere with the radius $R$
\begin{equation}
D\equiv R\operatorname{arcosh}[\sin y_0 \sin y_f +\cos y_0 \cos y_f \cos (x_f - x_0)].
\label{drift}
\end{equation}
The Lagrangian indicators can be computed by solving advection
equations forward and backward in time in order to know the fate and
origin of water masses, respectively. They have been shown recently
to be useful in quantifying transport of radionuclides in the
Northern Pacific after the accident at the Fukushima Nuclear Power
Plant \cite{DAN11,P13,NPG14} and in identifying the Lagrangian
Fronts favorable for fishing grounds \cite{DAN12,P13,DSR14}.
\begin{figure}[!htb]\center
\includegraphics[width=0.9\textwidth,clip]{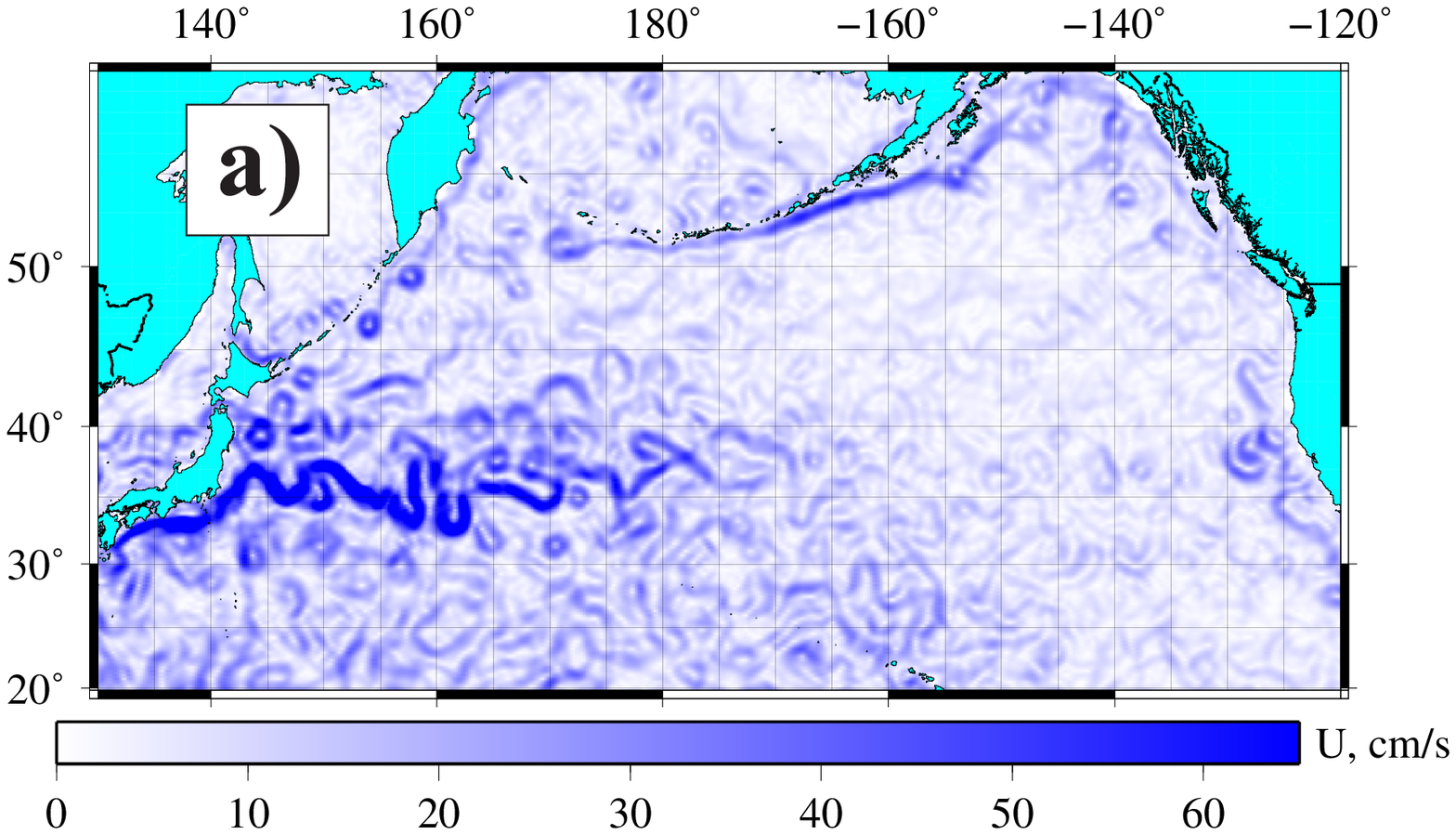}\\
\includegraphics[width=0.9\textwidth,clip]{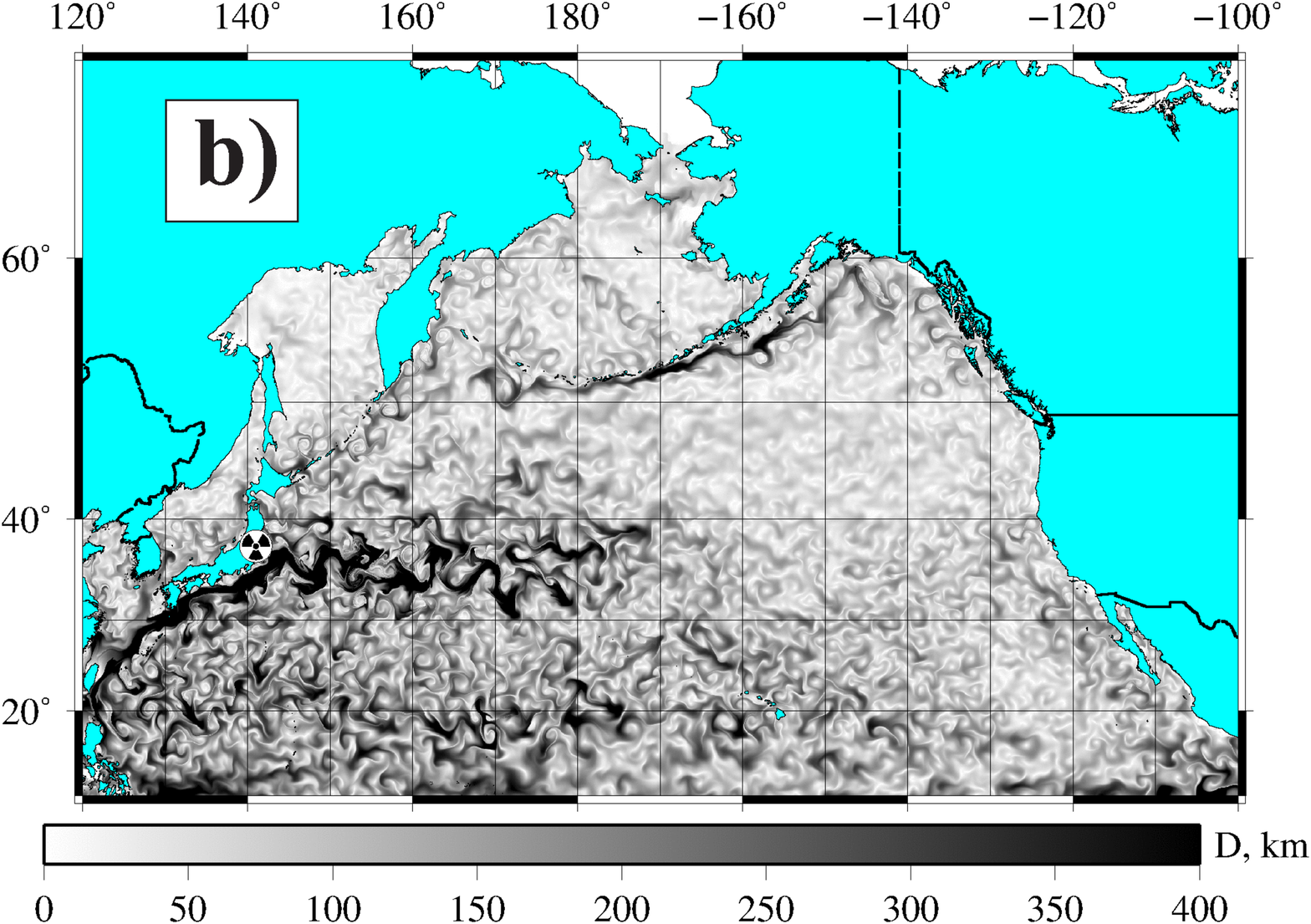}
\caption{a) Altimetric velocity field on 14 November 2010 ($U$ in cm/s) and 
b) the drift map ($D$ in km) for the North Pacific Ocean on 15 May 2011 
computed backward in time for two weeks.}
\label{fig3}
\end{figure}

Satellite-derived geostrophic velocity field on the fixed day, 14 November 2010, is shown
in Fig.~\ref{fig3}a for the enlarged area in the North Pacific Ocean.
The two powerful currents with increased speed values are visible on the map:
the Kuroshio and its Extension to the east of the Japan coast and the
Alaskan Stream along the Aleut Islands between the Kamchatka and Alaska peninsulas.

In Fig.~\ref{fig3}b we show the Lagrangian drift map on 15 May 2011 visualizing the absolute
displacements, $D$, for 2.25 millions of particles in the North Pacific
to be computed backward in time for two weeks in the altimetric AVISO velocity field.
The black color means that the corresponding water parcels on the map drifted considerably
as compared to the white colored particles.
Practically all the region is covered by  mesoscale eddies
of different sizes and dipole and mushroom-like structures. A few currents,
the Kamchatka, the Oyashio and the Californian ones, look like vortex streets with moving
mesoscale eddies each of which is surrounded by a black collar which
demarcates the boundary separating the eddy's core from the surrounding waters.

Color contrast on the drift maps demarcate boundaries between waters which
passed rather different distances before converging.
The map in Fig.~\ref{fig3}a demonstrates the ocean fronts on the planetary and synoptic
scales, including the subarctic frontal zone in the Japan Sea (situated between the Asia continent 
and Japan)
and low-energetic regions, as for example, the Okhotsk Sea (to the north of the Japan Sea)
excepting for its southern part between the Sakhalin and Kuril Islands.

The notion of a Lagrangian Front (LF), introduced recently \cite{P13,FAO14},
is defined as the boundary between waters with different Lagrangian properties.
It may be, for example, a physical property, such as
temperature, salinity, density, etc. or concentration of chlorophyll-$\alpha$.
Lateral maximal gradients of those properties would indicate on
specific oceanic fronts (thermal, salinity, density and chlorophyll ones)
which are often connected with each other. However, one may consider more
specific Lagrangian indicators such as absolute, meridional or zonal displacements of particles
from their initial positions and others. Even in the situation where the water itself
is indistinguishable, say, in temperature, and the corresponding sea-surface-temperature image does
not show a thermal front there may exist a LF separating waters with the other
distinct properties \cite{P13,FAO14}.
\begin{figure}[!htb]\center
\includegraphics[width=0.49\textwidth,clip]{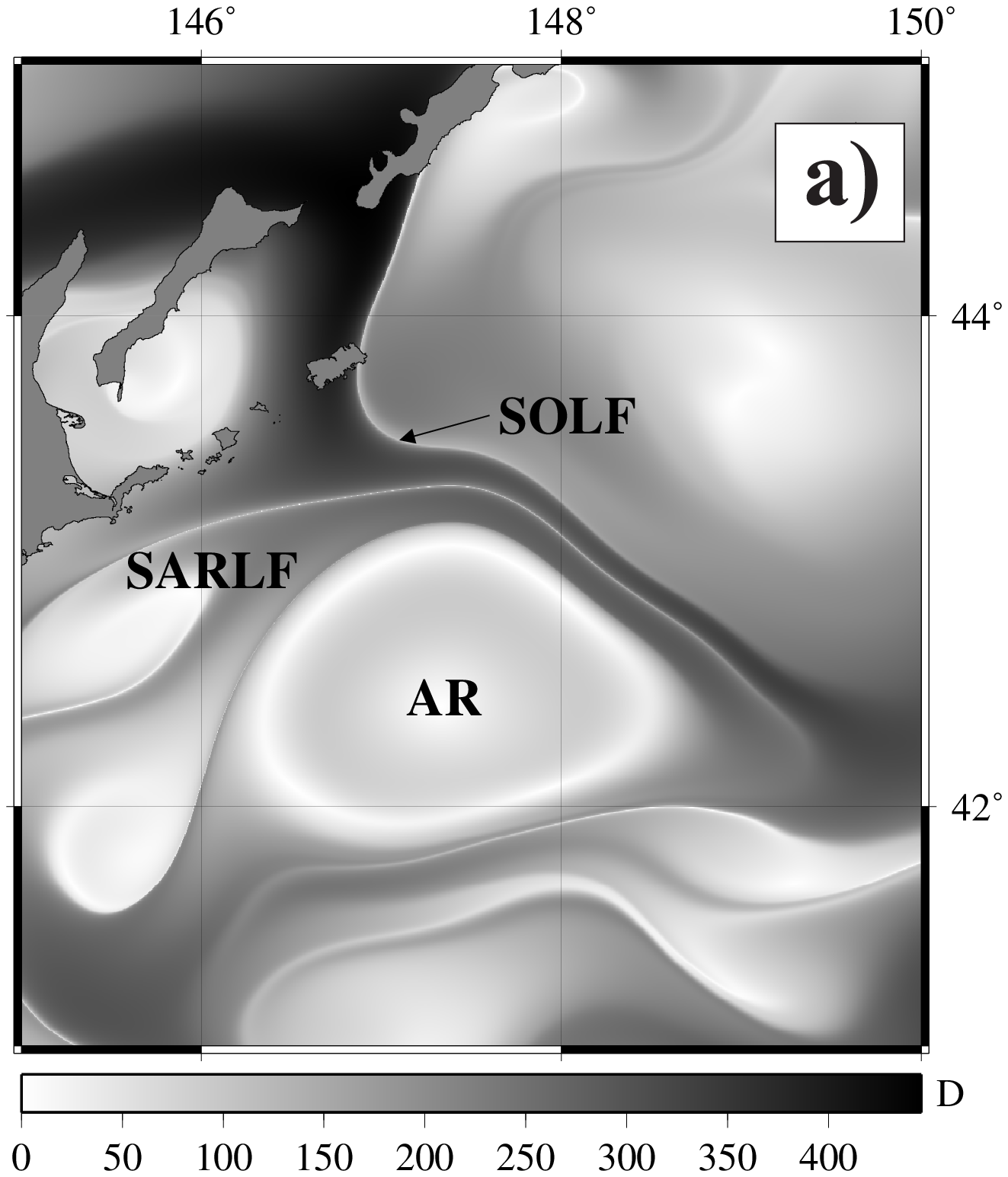}
\includegraphics[width=0.49\textwidth,clip]{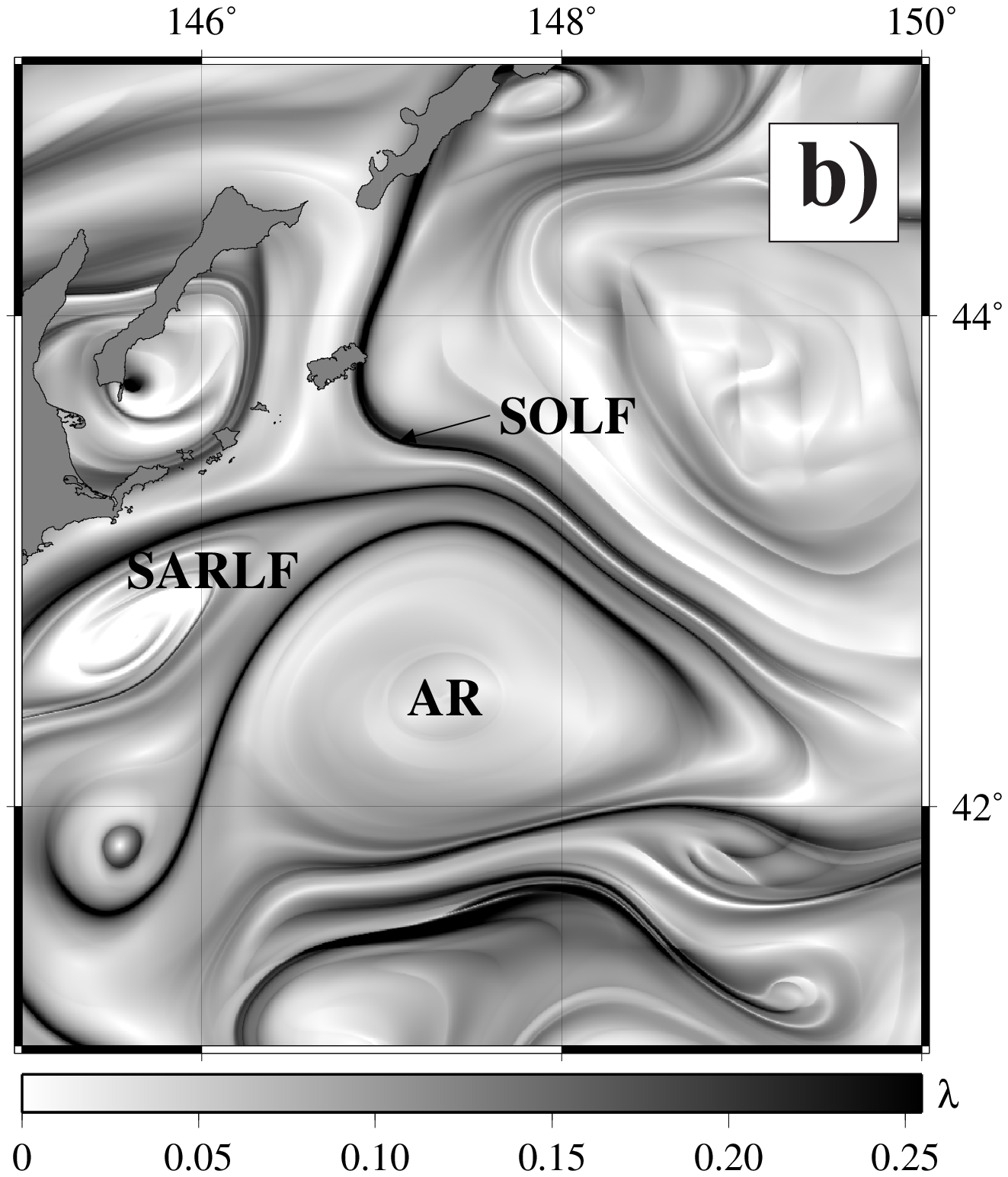}
\caption{a) Drift and b) Lyapunov maps in the region to the east off the Hokkaiso Island
and the Kuril Islands on 1 October 2004 computed backward in time for two weeks.}
\label{fig4}
\end{figure}
In the satellite era, it becomes possible to monitor common thermal and chlorophyll fronts
on the images of sea surface temperature and color, respectively.
The definition of the LF given above describes any frontal feature even the one
where similar waters from different places converge.
For example, one may code by different colors the synthetic particles,
that enter into the area under study through different geographical borders. The border between
the distinct colors on the corresponding Lagrangian map would be a kind of the LF
no matter how different are the properties of convergent waters. However,
in practice any LF demarcates convergence of dissimilar waters.

Restricting the area under study, one could resolve fine Lagrangian structures.
As an example we plot in Fig.~\ref{fig4} the drift and FTLE maps for the region to the east
off the Hokkaido Island and the southern Kuril Islands.
A large number of synthetic particles were distributed over that region and integrated
backward in time for two weeks to compute the absolute
particle's displacements from their initial positions.
Coding the particle's displacements by color,
one gets information on origin and history of water
masses present in the region on a given day.
The main regional LFs become visible on the Lagrangian map in Fig.~\ref{fig4}a.
The SOLF visualizes a convergence of Oyashio waters with
the Soya Current flowing from the west through the straits between the islands.  
The SELF separates the Soya waters from the anticyclonic Kuroshio ring ones
with the center at ($147^{\circ}.5E; 42^{\circ}.4 N$).
Each of the LFs can be identified by a narrow white band demarcating
the curve of the maximal gradient of $D$. White color
means that the corresponding particles have experienced very small displacements
over two weeks. In order to understand why it is so, we have computed the FTLE in the region.
The map in Fig.~\ref{fig4}b demonstrates clearly
the Kuroshio ring surrounded by black ridges which are known to approximate
unstable manifolds of the HTs in the region. The black ridges
in Fig.~\ref{fig4}b are situated along the corresponding white curves with
maximal gradients of $D$ because motion of particles nearby HPs 
slows down due to the presence of so-called saddle dynamical traps \cite{Chaos07}.

It should be stressed that the Lagrangian maps provide a new information on the flow structure
and its history that one cannot get looking at altimetric velocity-field daily snapshots
or satellite images of sea surface temperature, because the field may (and it really does)
fluctuate rather erratically, and the temperature images are not available during cloudy and rainy days.
Moreover, those maps enable us to compute ``exact'' positions of Largangian fronts and LCSs
in the region.

One of the practical applications of the Lagrangian analysis
has been provided recently in \cite{DAN12,DSR14}.
Using Pacific saury catch and location data of the Russian Fishery Agency
for a number of commercial fishery seasons, it was shown statistically
that the saury
fishing grounds with maximal catches are not randomly distributed over the region but located mainly
along the sharp regional LFs where productive cold waters of the Oyashio Current, warmer waters
of the southern branch of the Soya Current and waters of
warm-core Kuroshio rings converge. Possible biophysical reasons for accumulation of saury schools
at some major LFs in the region have been discussed in Refs.~\cite{P13,DSR14}.
The impact of LCSs on biological organisms has been studied in Refs.~\cite{Kai09,Ovidio13}.
By comparing the
seabird satellite positions with computed LCSs locations,
it was found in Ref.~\cite{Kai09} that a top marine predator, the Great Frigatebird, was able
to track the LCSs
in the Mozambique Channel identified with the help of the FTLE field.
As to another species, they may prefer another ocean features to accumulate at. For example,
tuna does not tend
to aggregate for feeding at LFs. It is seen in Figs.~\ref{fig2}a and d that tuna catch locations
are not correlated with the LFs and the LCS in the region. Tuna rather prefers
to use mesoscale eddies to move along their boundaries to seek for a food.
The recent study \cite{Ovidio13} of the behavior of tagged elephant seals in the Southern Indian Ocean
has shown that they prefer to cross several eddies before staying for intensive searching only in
an eddy with large retention time.

Lagrangian fronts can be accurately detected in
a given velocity field by computing Lagrangian maps of displacements of synthetic
tracers and other Lagrangian indicators.
The question is how they correlate with the LCS is open up to now.
The Lagrangian indicator, specifying a LF, varies significantly on both sides of the LF.  
The FTLE values 
are almost the same on both sides of any ridge in the FTLE field. Local extrema of that field approximate 
the corresponding LCSs. By definition, any Lagrangian indicator 
is a function of trajectory, whereas in order to compute the FTLE it is necessary in addition 
to know the dynamical system as well.
Displacement and the other Lagrangian indicators are characteristics of a given fluid 
particle whereas the Lyapunov exponent is a characteristic of the medium surrounding that particle.
We would like to stress the important role of LFs because,
in difference from rather abstract geometric objects of an associated dynamical system, like stable and unstable invariant manifolds, 
they are fronts of real physical quantities that can be, in principle, measured directly.

\section{Tracking Fukushima-derived radionuclides}
\label{sec:6}

The material line technique developed in Ref.~\cite{NPG14}
is a tool to trace origin, history and fate of water masses.
The material line with a large number of particles (markers),
crossing a feature under study, evolves backward in time.
It is useful sometimes to get as an output tracking Lagrangian maps
showing by density plots where the corresponding
markers were walking for a given period of time.
Placing markers inside a specific mesoscale eddy
along the transects where {\em in-situ} measurements have been carried out before,
we can simulate the history and origin of that eddy. It has been done in
\cite{NPG14} with Kuroshio rings, large cyclonic and anticyclonic eddies to be pinched off
from the meandering Kuroshio Current. The corresponding tracking and drift maps have allowed to document
near-surface transport of water masses across the strong Kuroshio Extention jet.
That simulation results were supported by tracks of the surface drifters which were deployed
in the area \cite{NPG14}.

The tracking technique may be useful in planning research vesseal cruises in the ocean.
Before choosing the track of a planed cruise, it is instructive to make a simulation by
initializing backward-in-time evolution of
material lines, crossing potentially interesting coherent structures in the
region visible on Lagrangian maps. The corresponding tracking maps would help us to know where
one could expect, for example, higher or lower concentrations of radionuclides, pollutants or other Lagrangian
tracers. This idea has been used in the ``Professor Gagarinskiy'' cruise to be conducted
in the area shown in Fig.~\ref{fig2} from 12 June to 10 July 2012, 15 months after the
accident at the Fukushima Nuclear Power Plant on 11 March 2011 \cite{Lobanov}.
Large amount of water contaminated with radionuclides leaked directly
into the ocean. Moreover, the radioactive pollution of the sea after the accident was caused
by atmospheric deposition on the ocean surface.
Just after the accident radioactive $^{137}{\rm Cs}$ and $^{134}{\rm Cs}$ with 30.07 yrs and 2.07 yrs
half-lifes, respectively, have been detected over a broad area in the North
Pacific. Before March 2011, $^{137}$Cs concentration
levels off Japan were 1 -- 2 Bq~m$^{-3}$ $\simeq$
0.001 -- 0.002 Bq~kg$^{-1}$ while $^{134}{\rm Cs}$ was not detectable.
Because of a comparatively short half-life time, any measured concentrations of
$^{134}$Cs could only be Fukushima derived.
\begin{figure}[!htb]\center
\includegraphics[width=0.7\textwidth,clip]{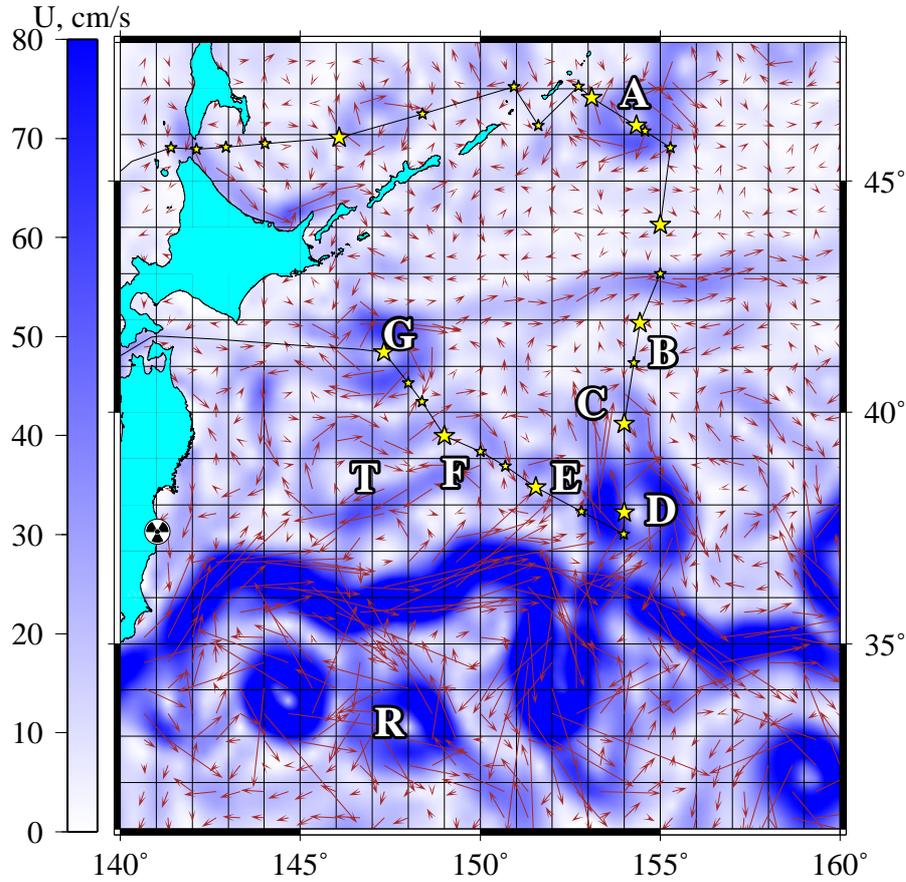}
\caption{Altimetric velocity field on 28 June 2012  
around the Fukushima Nuclear Power Plant (the radioactivity sign)  
with anticyclonic mesoscale eddies A, B, C, D, E, F and G
where seawater samples have been collected in the cruise \cite{Lobanov} in
the end of June and beginning of July 2012. The ship's track and some sampling stations are shown.}
\label{fig5}
\end{figure}
\begin{figure}[!htb]\center
\includegraphics[width=0.49\textwidth,clip]{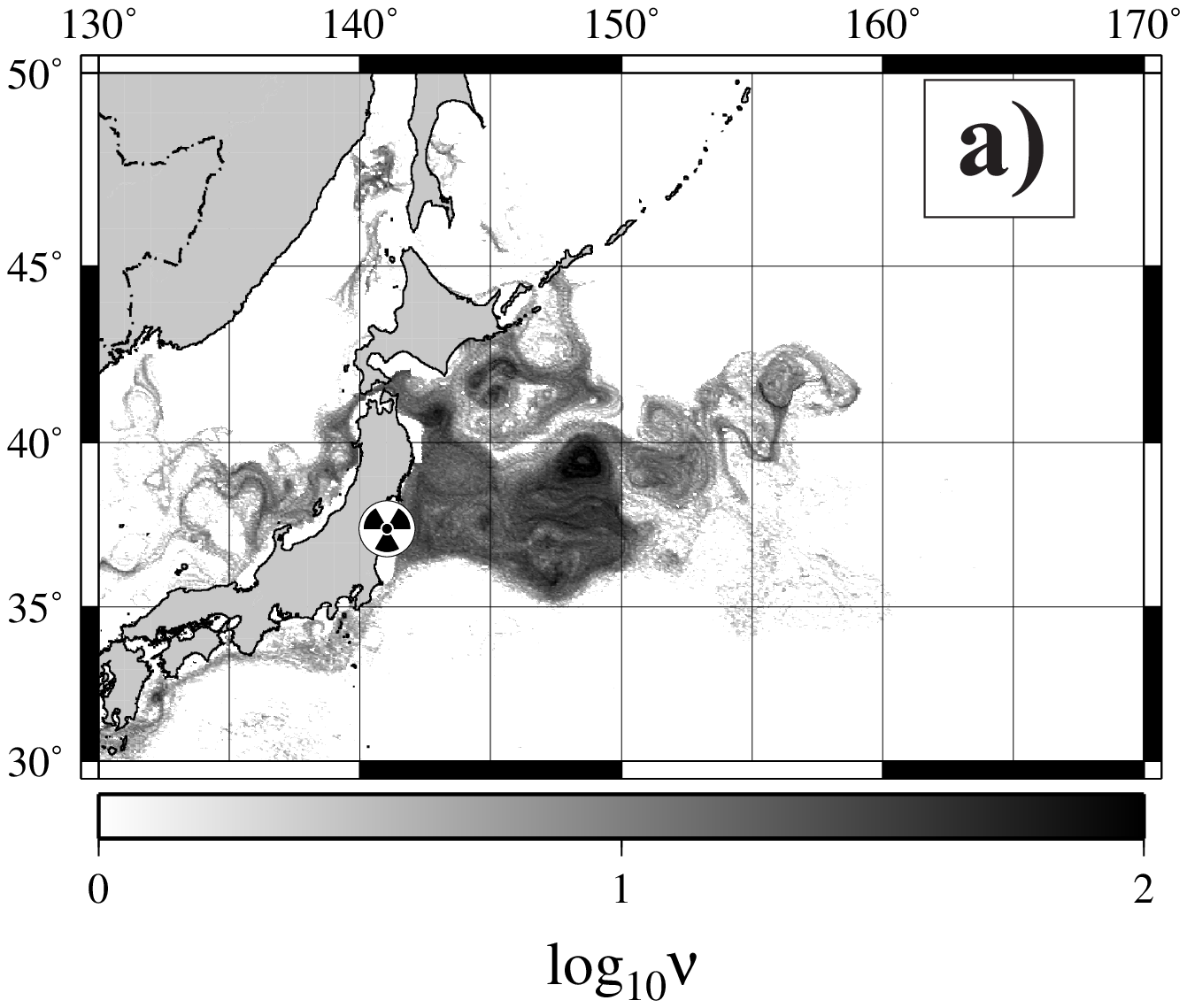}
\includegraphics[width=0.49\textwidth,clip]{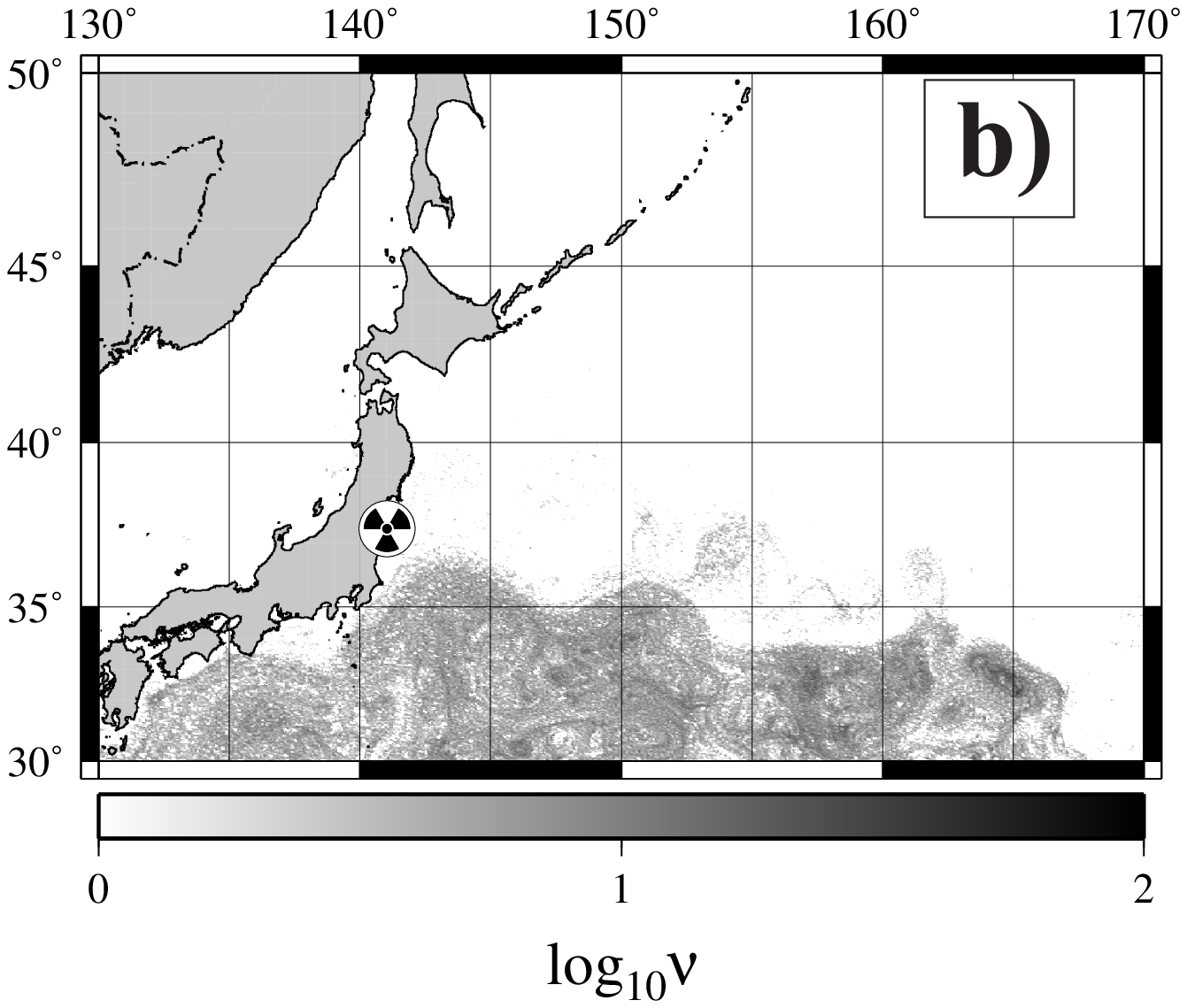}\\
\includegraphics[width=0.49\textwidth,clip]{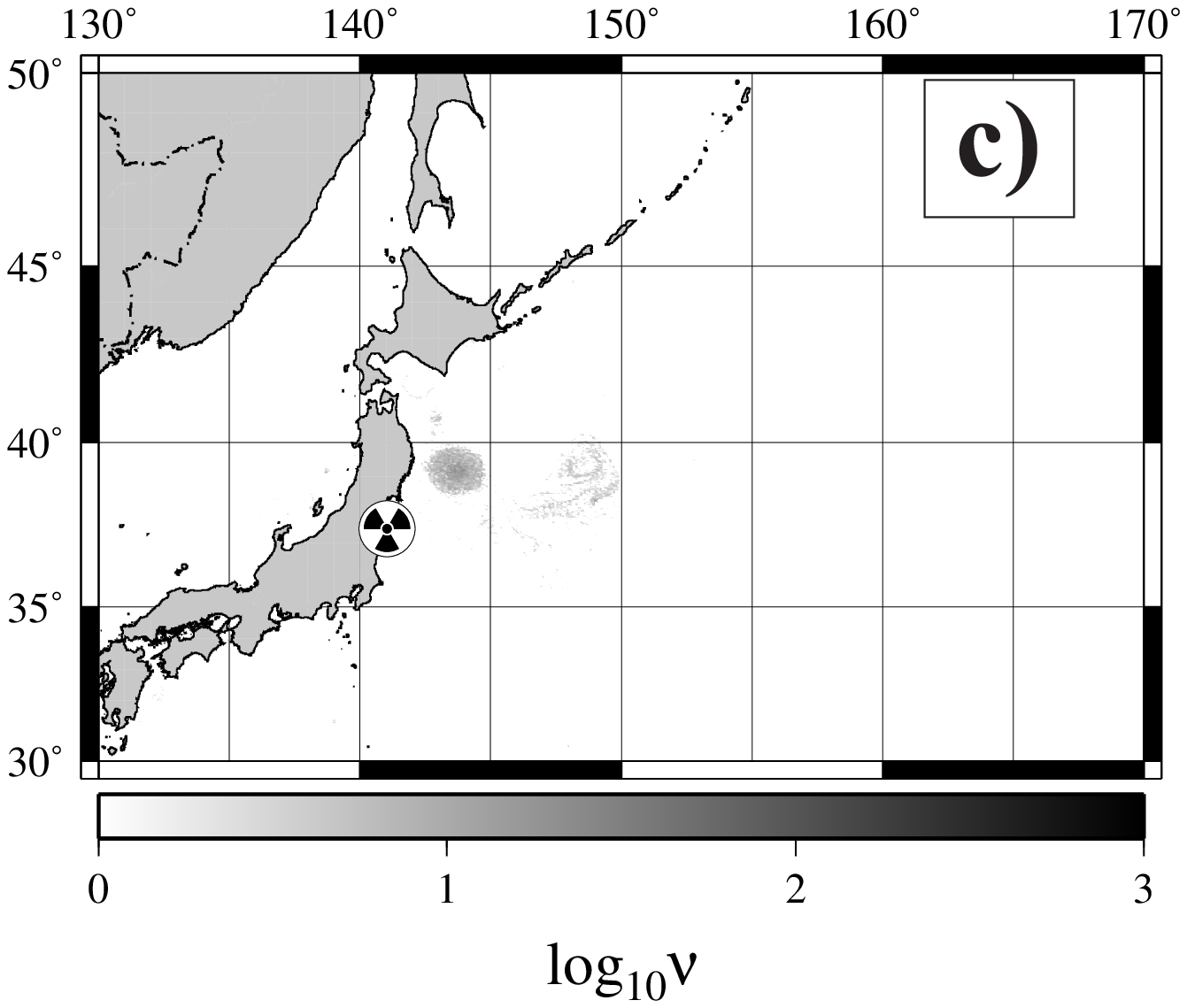}
\includegraphics[width=0.49\textwidth,clip]{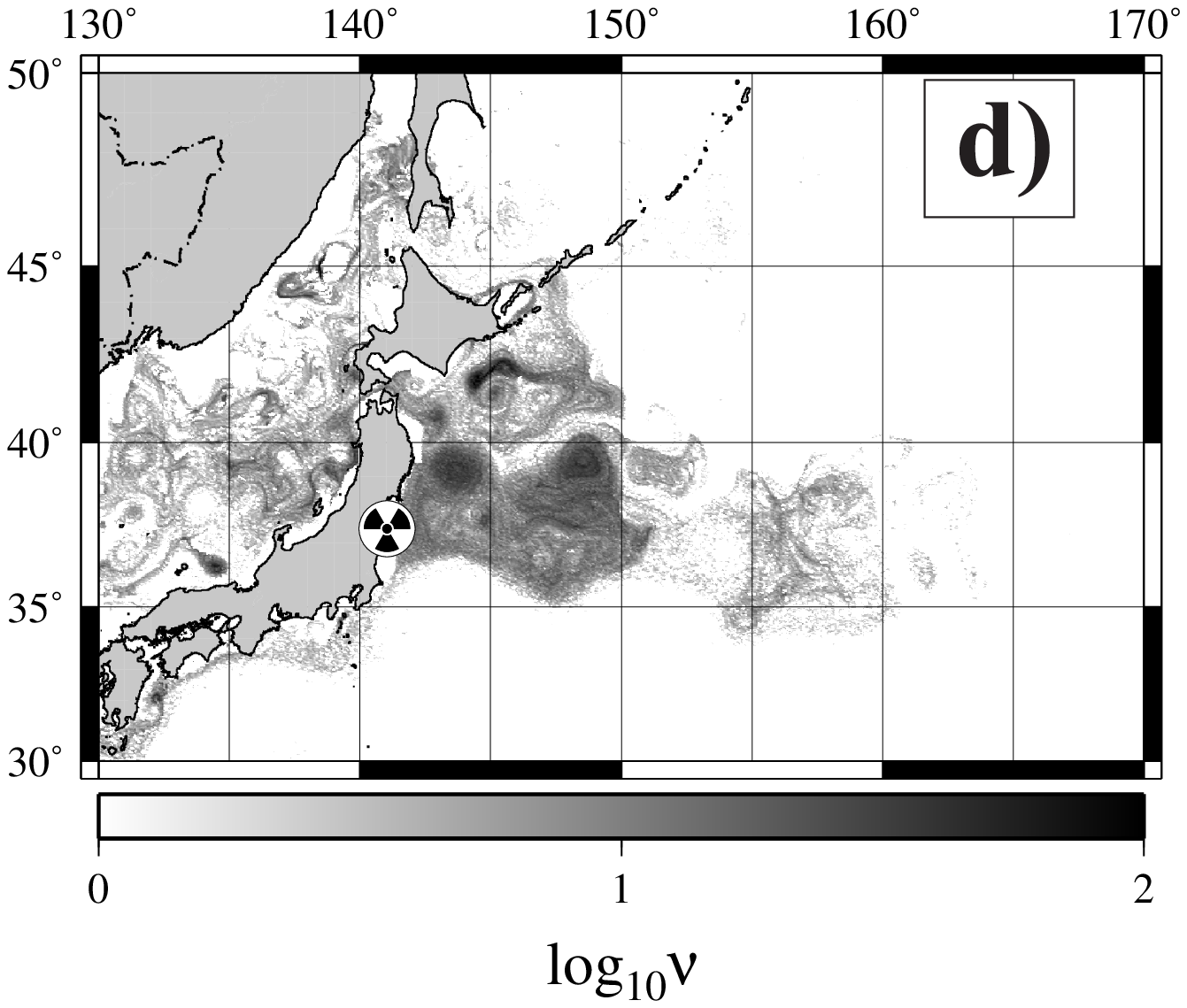}
\caption{Simulated tracking maps for the synthetic tracers distributed
in the centers of the eddies a) B, b) D, c) E and d) G.  
The maps show by the density plots where those tracers were walking during the month after the accident.
The density of traces, $\nu$, is in a logarithmic scale.}
\label{fig6}
\end{figure}

Lagrangian drift and FTLE maps, based on AVISO-provided altimeter
velocity fields, have been computed daily and sent by email on the
board. The researchers could see daily state of the ocean surface
with eddies, strong jets and streamers and plan the cruise track and
positions of sampling stations. The frontal Kuroshio -- Oyashio zone
is populated with mesoscale eddies of different sizes and lifetimes (Fig.~\ref{fig5}).
The cruise track, shown on the altimetric velocity-field map in Fig.~\ref{fig5},  
was chosen to cross the anticyclonic eddies A, B, C, D, E, F and G as perspective 
features for measuring caesium concentration in seawater samples collected at
different depth horizons. In simulation, a region around a sampling
station was populated with synthetic tracers which have been advected in an
altimetric velocity field backward in time beginning from the date
of sampling to the day of the accident. Fixing the places on a
tracking map, where the corresponding tracers were walking for one
month after the accident, the authors \cite{Lobanov} were able to
estimate by the trace density the probability to detect an increased
concentration of Fukushima-derived radionuclides in surface seawater
samples at a given station.

The results of simulation have been in a good agreement with {\em in situ} measurements
of $^{134}$Cs and $^{137}$Cs.
Four tracking maps are shown in Fig.~\ref{fig6} where density of traces, $\nu$, is shown 
in a logarithmic scale. 
Station 61 ($154^{\circ}.4E; 41^{\circ}.9 N$) was located
near the elliptic point of the eddy B with the size of $\simeq 1.5^{\circ}N \times 1.5^{\circ}E$.
The highest caesium concentrations, $21.1 \pm 1.1$ at surface and
$21.6 \pm 0.9$~Bq~m$^{-3}$ at 203~m depth,
have been observed in seawater samples at that station. It is well
agreed with measurements in another cruise \cite{Kaeriyama13} to be carried out
approximately in the same place at the same time.
They detected the concentration of $^{137}$Cs in surface seawater samples to be
$18 \pm 0.7$~Bq~m$^{-3}$
at their station B38 located nearby our station 61 and
$17 \pm 0.7$~Bq~m$^{-3}$ and
$13 \pm 0.7$~Bq~m$^{-3}$ at stations B37 and B39 located inside the eddy B.
The tracking map in Fig.~\ref{fig6}a shows that synthetic tracers, placed inside a patch around station 61,
have visited for the month
after the accident the places with presumedly high level of contamination.
In particular, they have often visited in March and April
2011 the Tohoku eddy (the eddy T in Fig.~\ref{fig5}) with the highest levels
of caesium concentration to be detected just after the accident \cite{Kaeriyama13}.
The place of the Tohoku eddy during the month after the accident is clearly seen in Fig.~\ref{fig6}c 
as a circular patch with increased density of points.
Tracing out the history of the eddy B \cite{Lobanov}, it was found that it was born on the southern
flank of a zonal eastward jet transporting waters from the eastern coast of the Honshu Island to
the open ocean.

The Kuroshio ring D with the size of $\simeq 3^{\circ}N \times
2.5^{\circ}E$ was pinched off from a meander of the Kuroshio
Extention in the end of May 2012. Until the middle of August it was
a free ring to be connected sometimes with the parent jet by an
arch. The probability to detect higher concentrations of caesium in
its core waters is estimated to be low (see Fig.~\ref{fig6}b)
because it contains mainly clean Kuroshio waters coming from the
south. The concentration of $^{137}$Cs, $6.3 \pm 0.4$~Bq~m$^{-3}$,
have been detected at station 69 in surface water samples
\cite{Lobanov}. It is slightly greater than the background level
that can be explained by water exchange with its companion, the eddy
C with higher level of radioactivity. The concentrations of
$^{137}$Cs in surface seawater samples at Japanese stations B30
located closely to station 69 have been found in \cite{Kaeriyama13}
to be close to the background level, $3.6 \pm 0.5$.

The Kuroshio ring E with the size of $\simeq 1^{\circ}N \times 1.5^{\circ}E$ was pinched off
from a meander of the jet on 10 -- 12 June 2012 and disappeared in the middle of July.
Station 74 ($151^{\circ}.5 E; 38^{\circ}.38 N$) was located
near the elliptic point of that eddy where the increased concentration
of $^{137}$Cs, $12.3 \pm 0.8$~Bq~m$^{-3}$, has been detected at 307~m depth \cite{Lobanov}.
A comparatively small number of traces over the whole broad area in Fig.~\ref{fig6}c
is explained by the history of its core water which have been transported
mainly by the Kuroshio from the south and then directed to the east by the
Kuroshio Extention. The genesis of the eddy E has shown a presence of the Tohoky eddy waters in its core
(see the patch in Fig.~\ref{fig6}c).

Station 84 ($147^{\circ}.3 E; 41^{\circ}.3 N$)
was located near the elliptic point of the eddy G with the size of $\simeq 1^{\circ}.5 N \times 2^{\circ}E$
situated at the traverse of the Tsugaru Strait between the Honshu and Hokkaido islands. 
The tracking map for
that station in Fig.~\ref{fig6}d reveals
its close connection with the Tohoku eddy, and, therefore, the probability to detect increased
caesium concentrations  was expected to be comparatively large. In reality we detected the
concentration of $^{137}$Cs at 100~m depth to be $18 \pm 1.3$ ~Bq~m$^{-3}$.

{\it In situ} observations 15 months after the incident were compared with the results of simulation of
advection of Fukushima-derived cesium radioisotopes by the AVISO altimetric velocity field. The
computed tracking Lagrangian maps were used to reconstruct the history and origin of synthetic
tracers imitating measured seawater samples collected in the centers of some mesoscale anticyclonic eddies in
the North Western  Pacific. Increased Fukushima-derived cesium-isotopes concentrations have been
detected in June and July 2012 at sampling stations located in the centers of anticyclonic eddies
B and G whose core waters have been demonstrated numerically to visit the areas with presumably high level of
contamination just after the accident (Figs.~\ref{fig6}a and d). Fast water advection between
anticyclonic eddies and convergence of
surface water inside eddies make them responsible for spreading and accumulation of cesium rich water.

\section{Conclusion}
\label{sec:7}
The dynamical systems approach provides a new way for describing large-scale chaotic transport and mixing
in geophysical flows. It uses tools from dynamical chaos theory to
find out and quantify organizing Lagrangian structures in a flow and their impact on large-scale motions.
The approach has been firstly applied to simplified analytic models of geophysical
flows and then to sophisticated numerical models of ocean circulation.
The research in this field is now shifting to study real ocean flows using
remote sensing data available to researchers from satellite observations, very high-frequency radars
and drifter deployments.

In this article we reviewed briefly some recent advances in applications of the
dynamical systems approach to the real ocean with the focus on synoptic maps of
some Lagrangian indicators. The basic theoretical ideas from chaotic advection
theory have been introduced firstly. Then we described briefly the Lagrangian approach to study
transport and mixing in the ocean and some numerical methods used to compute trajectories of
fluid particles in satellite-derived velocity fields, Lyapunov exponents and different
Lagrangian indicators. The Lagrangian maps, computed with a large number of synthetic tracers in
an area under study, have been shown to be useful in identifying the boundaries of different
regimes in that area including Lagrangian coherent structures and Lagrangian fronts.
We gave a number of illustrations of that approach.
As to practical applications, we discussed recent results by the present author and his
co-workers on identifying Lagrangian fronts favorable for fishing grounds and on tracking
the eddies in the North Pacific with increased concentration of Fukushima-derived radionuclides.

Research in this field has been very active in the last two decades. They seem to be
perspective in the future. Among the possible perspective ways of research we would like to mention 
the following ones. 1) Development of efficient Lagrangian methods for 3D flows.
2) Study of the role of Lagrangian fronts in behavior of marine organisms.
3) Analysis of impact of Lagrangian structures on drifter motion.

\section*{Acknowledgments}
I am grateful to Maxim Budyansky, Veniamin Razdobutko and Michael Uleysky
for their help in preparing some figures. This work was supported  by the Russian
Foundation for Basic Research
(project nos. 13-05-00099-a and 13-01-12404-ofi-m)
and by the Program ``Fundamental Problems of  Nonlinear Dynamics in Mathematical
and Physical Sciences'' of the Russian Academy of Sciences.


\begin{thebibliography}{}
%
\bibitem{A65}  V.I. Arnold, C. R. Hebd. Seances Acad. Sci. {\bf 261}, 17 (1965).
%
\bibitem{H66}  M. Henon,  C. R. Hebd. Seances Acad. Sci. {\bf 262}, 312 (1966).
%
\bibitem{Ar84} H. Aref, J. Fluid Mech. {\bf 143}, 1 (1984).
%
\bibitem{Ar02} H. Aref, Phys. Fluids {\bf 14}, 1315 (2002).
%
\bibitem{Ottino} J.M. Ottino, {\em The kinematics of mixing: stretching, chaos, and transport}
(Cambridge University Press, Cambridge, 1989).
%
\bibitem{PhysD04} M. Budyansky, M. Uleysky, S. Prants, Physica D {\bf 195}, 369 (2004).
%
\bibitem{JETP04} M.V. Budyansky, M.Yu. Uleysky, S.V. Prants, Journal of Experimental and
Theoretical Physics {\bf 99}, 1018 (2004).
%
\bibitem{Wiggins05} S. Wiggins, Annu. Rev. Fluid Mech. {\bf 37}, 295 (2005).
%
\bibitem{KP06} K.V. Koshel, S.V. Prants, Physics -- Uspekhi {\bf 49}, 1151 (2006).
%
\bibitem{Zaslavsky} S. Abdullaev, G. Zaslavsky, Usp. Fiz. Nauk {\bf 8} 1 (1991).
%
\bibitem{Chaos04} D.V. Makarov, M.Yu. Uleysky, S.V. Prants, Chaos {\bf 14}, 79 (2004).
%
\bibitem{Makarov} D. Makarov, S. Prants, A.~ Virovlyansky,
G.~ Zaslavsky, {\em Ray and wave chaos in ocean acoustics: chaos in waveguides}
(World Scientific, Singapore, 2010).
%
\bibitem{UFN12} A.L. Virovlyansky, D.V. Makarov, S.V. Prants,
Physics--Uspekhi {\bf 55}, 18 (2012).
%
\bibitem{S92} R.M Samelson, J. Phys. Oceanogr. {\bf 22}, 431 (1992).
%
\bibitem{Chaos06} S.V. Prants, M.V. Budyansky, M.Yu. Uleysky, G. M. Zaslavsky,
Chaos {\bf 16}, 033117 (2006).
%
\bibitem{Chaos07} M.Yu. Uleysky, M.V. Budyansky, S.V. Prants, Chaos {\bf 17}, 024703 (2007).
%
\bibitem{JPA08} M.Yu. Uleysky, M.V. Budyansky, S.V. Prants, J. Phys. A {\bf 41}, 215102 (2008).
%
\bibitem{PRE09} M.V. Budyansky, M.Yu. Uleysky, S.V. Prants, Phys. Rev. E {\bf 79}, 056215 (2009).
%
\bibitem{P91} R.T. Pierrehumbert, Geophys. Astrophys. Fluid Dyn. {\bf 58}, 285  (1991).
%
\bibitem{P94} R.T. Pierrehumbert, Chaos, Solit. Fract. {\bf 4}, 1091  (1994).
%
\bibitem{DM93} D. Del-Castillo-Negrete, P.J. Morrison, Phys. Fluids A {\bf 5} 948  (1993).
%
\bibitem{Kozlov99} V.F. Kozlov, K.V. Koshel, Izvestiya Akademii Nauk
Fizika Atmosferi i Okeana {\bf 35}, 137 (1999).
%
\bibitem{Rypina}  I.I. Rypina,  M.G. Brown, F.J. Beron-Vera, H. Kozak,
M.J. Olascoaga, I.A. Udovydchenkov, J. Atmos. Sci. {\bf 64}, 3595 (2007).
%
\bibitem{Koshel08} K.V. Koshel, M.A. Sokolovskiy, P.A. Davies, Fluid Dyn. Res. {\bf 40}, 695 (2008).
%
\bibitem{Koshel10} E.A. Ryzhov, K.V. Koshel, D.V. Stepanov, Theor. Comput. Fluid Dyn. {\bf 24}, 59 (2010).
%
\bibitem{PRE10} M.Yu. Uleysky, M.V. Budyansky, S.V. Prants, Phys. Rev. E {\bf 81}, 017202 (2010).
%
\bibitem{JETP10} M.Yu. Uleysky, M.V. Budyansky, S.V. Prants, Journal of Experimental and
Theoretical Physics {\bf 111}, 1039 (2010).
%
\bibitem{Zhmur11} V.V. Zhmur, E.A. Ryzhov, K.V. Koshel, J. Mar. Res. {\bf 69}, 435 (2011).
%
\bibitem{Sok11} M.A. Sokolovskiy, K.V. Koshel, X. Carton,  Geophys. Astrophys. Fluid Dyn. {\bf 105},
505 (2011).
%
\bibitem{Ryz11} E.A. Ryzhov, K.V. Koshel, Izvestiya, Atmospheric and Oceanic Physics {\bf 47},
241   (2011)
%
\bibitem{Koshel13}  K.V. Koshel, M.A. Sokolovskiy, J. Verron,  J. Fluid Mech. {\bf 717}, 255 (2013).
%
\bibitem{SMS89} J. Sommeria, S.D. Meyers, H.L. Swinney, Nature {\bf 337}, 58 (1989).
%
\bibitem{SHS93}  T.H. Solomon, W.J. Holloway, H.L. Swinney, Phys. Fluids A {\bf 5}, 1971  (1993).
%
\bibitem{H02} G. Haller, Phys. Fluids {\bf 14}, 1851 (2002).
%
\bibitem{Hussain83} A.K.M.F. Hussain, Phys. Fluids {\bf 26}, 2816  (1983).
%
\bibitem{Haller13} T. Peacock, G. Haller,  Phys. Today {\bf 66 (2)}, 41 (2013).
%
\bibitem{Samelson13} R.M. Samelson,  Ann. Rev. Mar. Sci. {\bf 5}, 137 (2013).
%
\bibitem{Abraham02} E.R. Abraham, M.M. Bowen, Chaos {\bf 12}, 373 (2002).
%
\bibitem{Ovidio04} F. d'Ovidio, V. Fernandez, E. Hernandez-Garcia, C. Lopez,
Geophys. Res. Lett. {\bf 31}, L17203 (2004).
%
\bibitem{Shadden05} S. Shadden, F. Lekien, J.E. Marsden, Physica D {\bf 212}, 271 (2005).
%
\bibitem{Kirwan06} A.D. Jr Kirwan, Prog. Ocean. {\bf 70}, 448 (2006).
%
\bibitem{Lehahn07} Y. Lehahn, F. d'Ovidio, M. Levy, E. Heifetz, J. Geophys. Res. {\bf 112}, C08005 (2007).
%
\bibitem{Beron08} F. Beron-Vera, M. Olascoaga, G. Goni, Geophys. Res. Lett. {\bf 35}, L12603  (2008).
%
\bibitem{Kai09} E. Tew Kai, V. Rossi, J. Sudre, H. Weimerskirch, C. Lopez,
E. Hernandez-Garcia, F. Marsac, V. Garcon, Proc. Nat. Ac. Sci. USA {\bf 106}, 8245 (2009).
%
\bibitem{OM11} S.V. Prants, M.V. Budyansky, V.I. Ponomarev, M.Yu. Uleysky,
Ocean modelling {\bf 38}, 114 (2011).
%
\bibitem{DAN12} S.V. Prants, M.Yu. Uleysky, M.V. Budyansky, Doklady Earth Sciences {\bf 447},
1269 (2012).
%
\bibitem{FAO13} S.V. Prants, V.I. Ponomarev, M.V. Budyansky,  M.Yu. Uleysky, P.A. Fyman,
Izvestiya, Atmospheric and Oceanic Physics {\bf 49}, 82  (2013).
%
\bibitem{P13} S.V. Prants, Physica Scripta {\bf 87}, 038115 (2013).
%
\bibitem{OM13} S.V. Prants, A.G. Andreev, M.V. Budyansky, M.Yu. Uleysky.
Ocean Modelling {\bf 72}, 143 (2013).
%
\bibitem{DSR14} S.V. Prants, M.V. Budyansky, M.Yu. Uleysky, Deep Sea Res. I {\bf 90}, 27 (2014).
%
\bibitem{FAO14} S.V. Prants, M.V. Budyansky, M.Yu. Uleysky,
Izvestiya, Atmospheric and Oceanic Physics {\bf 50}, 284 (2014).
%
\bibitem{NPG14} S.V. Prants, M.V. Budyansky, M.Yu. Uleysky, Nonlinear Proc. Geophys. {\bf 21}, 279 (2014).
%
\bibitem{Lobanov} M.V. Budyansky, V.A. Goryachev, D.D. Kaplunenko, V.B. Lobanov, S.V. Prants,
A.F. Sergeev, N.V. Shlyk, M.Yu. Uleysky, Deep Sea Res. I (in press).
%
\bibitem{OD14} S.V. Prants, A.G. Andreev, M.Yu. Uleysky, M.V. Budyansky.
Ocean Dynamics {\bf 64}, 771 (2014).
%
\bibitem{DAN11} S.V. Prants, M.Yu. Uleysky, M.V. Budyansky, Doklady Earth Sciences {\bf 439},
1179 (2011).
%
\bibitem{Huhn12} F. Huhn F, A. von Kameke, V. Perez Munuzuri, M.J. Olascoaga,
F.J. Beron-Vera, Geophys. Res. Lett. {\bf 39}, L06602 (2012).
%
\bibitem{Bettencourt12} J.H. Bettencourt, C. Lopez, E. Hernandez-Garcia, Ocean Modelling
{\bf 51}, 73 (2012).
%
\bibitem{Beron13} F.J. Beron-Vera, Y. Wang, M.J. Olascoaga, G.J. Goni, G. Haller,
J. Phys. Oceanogr. {\bf 43}, 1426 (2013).
%
\bibitem{PRE06} D.V. Makarov, M.Yu.~Uleysky, M.V~Budyansky, S.V.~Prants,
Phys. Rev. E {\bf 73},  066210 (2006).
%
\bibitem{Wig06} A.M. Mancho, Des Small, S. Wiggins, Phys. Rep. {\bf 437}, 55 (2006).
%
\bibitem{Ovidio13} F. d'Ovidio, S. de Monti, A.D. Penna, C. Cotte, C. Guinet, J. Phys. A {\bf 46},
254023 (2013).
%
\bibitem{Kaeriyama13} H. Kaeriyama, et al, Biogeosciences {\bf 10}, 4287 (2013).
%
\end{thebibliography}
\end{document}